\documentclass[traditabstract]{aa}
\usepackage{natbib,graphicx}
\bibpunct{(}{)}{;}{a}{}{,} 
\usepackage{amsmath,amsfonts,amssymb}
\usepackage{txfonts}
\usepackage{longtable}
\usepackage{soul,color,lscape}
\usepackage[breaklinks,colorlinks,citecolor=blue]{hyperref}
\usepackage{float}
\usepackage{multicol}
\usepackage{stfloats}

\usepackage{todonotes,ifthen}
\newcounter{mycomment}
\newcommand{\mycomment}[2][]{%
    \refstepcounter{mycomment}%
    \ifthenelse{\equal{#1}{NM}}
    {\todo[color={red!40},size=\small, inline]{%
     \textbf{{#1}\themycomment:}~#2}%
    }{\ifthenelse{\equal{#1}{JHD}}
    	{\todo[color={blue!40},size=\small, inline]{%
     	\textbf{{#1}\themycomment:}~#2}%
    	}{\ifthenelse{\equal{#1}{EJ}}
        {\todo[color={green!40},size=\small, inline]{%
     	\textbf{{#1}\themycomment:}~#2}%
    	}{}}}\noindent}

\newcommand{\new}[1]{#1}

\begin{document}

\title{Probing dark energy with tomographic weak-lensing aperture mass statistics}

\titlerunning{Probing dark energy with aperture mass statistics}
\authorrunning{Martinet et al.}

\author{Nicolas Martinet$^{1}$, Joachim Harnois-D\'eraps$^{2,3}$, Eric Jullo$^{1}$, Peter Schneider$^{4}$}

\institute{$^{1}$ Aix-Marseille Univ, CNRS, CNES, LAM, Marseille, France\\
$^{2}$ Institute for Astronomy, University of Edinburgh, Royal Observatory, Blackford Hill, Edinburgh EH9 3HJ, UK\\
$^{3}$ Astrophysics Research Institute, Liverpool John Moores University, 146 Brownlow Hill, Liverpool, L3 5RF, UK\\
$^{4}$ Argelander-Institut  f\"ur  Astronomie,  Universit\"at  Bonn,  Auf  dem H\"ugel 71, 53121 Bonn, Germany\\
  \email{nicolas.martinet@lam.fr}}

\setcounter{page}{1}

\abstract{We forecast and optimize the cosmological power of various weak-lensing aperture mass ($M_{\rm ap}$) map statistics for future cosmic shear surveys, including peaks, voids, and the full distribution of pixels (1D $M_{\rm ap}$). These alternative methods probe the non-Gaussian regime of the matter distribution, adding complementary cosmological information to the classical two-point estimators. Based on the SLICS and cosmo-SLICS $N$-body simulations, we build \textit{Euclid}-like mocks to explore the $S_8 - \Omega_{\rm m} - w_0$ parameter space. We develop a new tomographic formalism that exploits the cross-information between redshift slices (cross-$M_{\rm ap}$) in addition to the information from individual slices (auto-$M_{\rm ap}$) probed in the standard approach. Our auto-$M_{\rm ap}$ forecast precision is in good agreement with the recent literature on weak-lensing peak statistics and is improved by $\sim 50$\% when including cross-$M_{\rm ap}$. It is further boosted by the use of 1D $M_{\rm ap}$ that outperforms all other estimators, including the shear two-point correlation function ($\gamma$-2PCF). When considering all tomographic terms, our uncertainty range on the structure growth parameter $S_8$ is enhanced by $\sim 45$\% (almost twice better) when combining 1D $M_{\rm ap}$ and the $\gamma$-2PCF compared to the $\gamma$-2PCF alone. We additionally measure the first combined forecasts on the dark energy equation of state $w_0$, finding a factor of three reduction in the statistical error compared to the $\gamma$-2PCF alone. This demonstrates that the complementary cosmological information explored by non-Gaussian $M_{\rm ap}$ map statistics not only offers the potential to improve the constraints on the recent $\sigma_8$--$\Omega_{\rm m}$ tension, but also constitutes an avenue to understanding the accelerated expansion of our Universe.}

\keywords{gravitational lensing: weak -- cosmology: observations -- surveys}

\maketitle


\section{Introduction}
\label{sec:intro}

The coherent distortion between galaxy shapes due to the gravitational lensing by large-scale structure has emerged as one of the most powerful cosmological probes today. So-called cosmic shear analyses have started to unlock their potential thanks to large observational surveys of hundreds of square degrees at optical wavelength: for example, CFHTLenS \citep{Heymans+12}, KiDS \citep{deJong+13}, DES \citep{Flaugher05}, and HSC \citep{Aihara+18}. In the meantime, the community is preparing for the next generation of cosmic shear surveys that will probe thousands of square degrees and extend to large redshifts: for example, \textit{Euclid} \citep{Laureijs+11}, Rubin Observatory \citep[formerly LSST,][]{Ivezic+19}, and Nancy Grace Roman Space Telescope \citep[formerly WFIRST,][]{Spergel+15}.

So far, these surveys have mostly relied on two-point estimators for their cosmological analysis, for example the shear two-point correlation functions \citep[$\gamma$-2PCF, e.g.,][]{Kilbinger+13, Troxel+18, Hikage+19, Asgari+20}. These estimators are inherited from cosmic microwave background (CMB) analyses, which probe the matter distribution of the early Universe. However, cosmic shear probes the recent Universe, where the matter distribution is more complex due to the nonlinear accretion of structures that creates non-Gaussian features in the matter field on small scales \citep[e.g.,][]{Codis+15,Barthelemy+20}. Two-point statistics fail to capture this non-Gaussian information and thus yield an incomplete description of the matter distribution at low redshift. To close this gap, the community has recently started to explore non-Gaussian cosmic shear estimators: for example weak-lensing peaks \citep[e.g.,][]{Kruse+99, Kruse+00, DH10, Kratochvil+10, Fan+10, Yang+11, Maturi+11, Hamana+12, Hilbert+12, Marian+12, Marian+13, Shan+14, Lin+15, Martinet+15b, Liu+15j, Liu+15x, Kacprzak+16, Petri+16, Zorilla+16, Shan+18, Martinet+18, Giocoli+18, Peel+18, Davies+19, Fong+19, Li+19, Weiss+19, Yuan+19, Coulton+20, Ajani+20, Zurcher+20}, Minkowski functionals \citep[e.g.,][]{Kratochvil+12, Petri+15, Vicinanza+19, Parroni+20, Zurcher+20}, higher-order moments \citep[e.g.,][]{vanWaerbeke+13, Petri+15, Peel+18, Vicinanza+18, Chang+18, Gatti+20}, three-point statistics \citep[e.g.,][]{Schneider+03, Takada+03, Takada+04, Semboloni+11, Fu+14}, density split statistics \citep[e.g.,][]{Friedrich+18, Gruen+18, Burger+20}, persistent homology \citep[e.g.,][]{Heydenreich+20}, scattering transform \citep[e.g.,][]{Cheng+20}, and machine learning \citep[e.g.,][]{Merten+19, Ribli+19, Peel+19, Shirasaki+19, Fluri+19}. Although these new statistics have shown a great potential, they have not yet been fully  generalized to data analyses, mostly because they need a large number of computationally expensive $N$-body simulations to calibrate their dependence on cosmology, while this dependence is accurately predicted by theoretical models in the case of two-point estimators.

In particular, weak-lensing peak statistics have proven to be one of the most powerful such new probes. Recent observation- \citep[e.g.,][]{Liu+15j, Kacprzak+16, Martinet+18} and simulation-based \citep[e.g.,][]{Zurcher+20} analyses show that peak statistics improve the constraints on \smash{$S_8=\sigma_8 \, \left(\Omega_{\rm m}/0.3\right)^{0.5}$} by $20$\% to $40$\% compared to the $\gamma$-2PCF alone. This particular combination of the matter density $\Omega_{\rm m}$ and clustering amplitude $\sigma_8$ runs across the main degeneracy direction obtained from cosmic shear constraints. It has notably been used to assess the potential tension between KiDS $\gamma$-2PCF cosmic shear and Planck CMB analyses \citep[e.g.,][]{Hildebrandt+17, Planck+18, Joudaki+20, Heymans+20}. In this article, we generalize the peak formalism to aperture-mass statistics, which provides additional constraining power on $S_8$.

 The aperture mass map ($M_{\rm ap}$ map in the following) is essentially a map of convergence, convolved with a ``Mexican hat''-like filter, which can be directly obtained from the estimated shear field (see Sect.~\ref{sec:map}). Other interesting $M_{\rm ap}$ map-sampling methods exist beside peaks: for example voids \citep[e.g.,][]{Gruen+16, Coulton+20} and the full distribution of pixels, also known as the lensing probability distribution function \citep[lensing PDF, e.g.,][]{Patton+17,Shirasaki+19, Liu+19}. Because they probe more massive structures, peaks are less sensitive to noise and are likely better for studying cosmological parameters sensitive to matter ($\Omega_{\rm m}$, $\sigma_8$) and the sum of neutrino mass \citep{Coulton+20}. Voids, however, could offer a better sensitivity to dark energy. The optimal choice of lensing estimator is currently unknown and needs to be determined.

In the context where Stage-IV weak lensing experiments are about to see first light, one of the key questions that needs to be addressed is how much improvement can we expect from higher-order statistics when it comes to constraining the dark energy equation of state $w_0$. In this paper, we present the first forecasts of this parameter from joint $M_{\rm ap}$ and $\gamma$-2PCF statistics and provide recommendations for applying tomography in a way that optimizes the total constraining power.

Indeed, while routinely used in $\gamma$-2PCF analyses, tomography has rarely been applied in mass map cosmological analyses \citep[e.g.,][]{Yang+11, Martinet+15b, Petri+16, Li+19, Coulton+20, Ajani+20, Zurcher+20}, less so with a realistic redshift distribution \citep[e.g.,][]{Martinet+15b, Petri+16, Zurcher+20}. One caveat of the current mass map tomography is that it only probes the information from individual redshift slices. This is similar to using only the auto-correlation terms in the case of the $\gamma$-2PCF and likely explains why the $\gamma$-2PCF benefits more from tomography than mass map estimators \citep{Zurcher+20}. In this article, we develop a new mass map tomographic approach to include the cross-information between redshift slices, and apply it to simulated lensing catalogs matched in depth to the future {\it Euclid} survey, with a broad redshift distribution between $0<z<3$.

The aim of this article is to serve as a basis for the application of $M_{\rm ap}$ non-Gaussian statistics to future cosmic shear surveys. In particular we review and optimize the statistical precision gain on $S_8$ and $w_0$ brought by different samplings of the $M_{\rm ap}$ map, either using the full distribution of pixels, the peaks, or the voids. We also optimize the tomographic approach in the context of a joint analysis with standard two-point estimators. We finally test the impact of performing the $M_{\rm ap}$ map computation in real or Fourier space, in presence of  observational masks, and study the effect of  parameter sampling explored with the $N$-body simulations.

The covariance matrix is built from the SLICS simulations \citep{Harnois-Deraps+15} while the cosmology dependence is modeled with the cosmo-SLICS  \citep{cosmoSLICS}, which sample the parameters $S_8$, $\Omega_{\rm m}$, $w_0$ and the reduced Hubble parameter $h$ at 25 points. This allows us to model the signals everywhere inside a broad parameter space and therefore to surpass the Fisher analyses often performed in the literature.

We first describe the mass map calculation that we use ($M_{\rm ap}$) in Sect.~\ref{sec:map}, and investigate its sensitivity to observational masks in Appendix~\ref{sec:res;subsec:mask}. We then present in Sect.~\ref{sec:sim} the SLICS  and cosmo-SLICS $N$-body simulation mocks. We detail the different data vectors (DVs) in Sect.~\ref{sec:vec} and our new tomographic approach in Sect.~\ref{sec:vec:subsec:tomo}. We present the likelihood pipeline used to make our cosmological forecasts in Sect.~\ref{sec:cos}, examining the cosmological parameter sampling in Appendix~\ref{sec:res;subsec:interp}. We forecast the constraints on $S_8$ and $w_0$ from $M_{\rm ap}$ statistics in Sect.~\ref{sec:res}, focusing on the tomography improvement (Sect.~\ref{sec:res;subsec:tomo}) and on the sampling of $M_{\rm ap}$ (Sect.~\ref{sec:res;subsec:samp}), and finally combine it with the $\gamma$-2PCF (Sect.~\ref{sec:res;subsec:2pcf}). We summarize our results in Sect.~\ref{sec:ccl}.

\section{Aperture mass computation}
\label{sec:map}

The $M_{\rm ap}$ statistic, first defined by \citet{Schneider96}, presents several advantages over classical mass reconstructions that particularly fit the purpose of cosmological analyses from mass maps. We first describe the general expression for $M_{\rm ap}$ in the shear and convergence space and discuss the pros and cons of this method (Sect.~\ref{subsec:Dir}). We then develop a Fourier space approach that increases the speed of the reconstruction (Sect.~\ref{subsec:Fou}).

\subsection{Real space computation}
\label{subsec:Dir}

The $M_{\rm ap}$ at a position $\smash{\vec{\theta}_0}$ consists in a convolution between the tangential ellipticity $\smash{\epsilon_{\rm t}(\vec{\theta},\vec{\theta}_0)}$ with respect to that point and a circularly symmetrical aperture filter $Q(\theta)$ (with $\theta = |\vec{\theta}|$),

\begin{equation}
\label{eq:map}
M_{\rm ap}(\vec{\theta}_0)=\frac{1}{n_{\rm gal}}\sum_i Q(|\vec{\theta}_i-\vec{\theta}_0|)\epsilon_{{\rm t}}(\vec{\theta}_i, \vec{\theta}_0),
\end{equation}

\noindent where the sum is carried out over observed galaxies and normalized by the galaxy density $n_{\rm gal}$ within the aperture. The tangential ellipticity is defined from the two components of the ellipticity and the polar angle $\smash{\phi(\vec{\theta},\vec{\theta}_0)}$ of the separation vector $\vec{\theta}-\vec{\theta}_0$ with respect to the center of the aperture $\smash{\vec{\theta}_0}$,

\begin{equation}
\label{eq:gt}
\epsilon_{\rm t}(\vec{\theta},\vec{\theta}_0)= - \Re\left[ \; \hat{\epsilon}(\vec{\theta}) \, {\rm e}^{-2{\rm i}\phi(\vec{\theta},\vec{\theta}_0)}\; \right].
\end{equation}

A hat denotes spin-2 quantities. Assuming that the mean intrinsic ellipticity within the aperture averages to 0, Eq.~(\ref{eq:map}) is an unbiased estimator of Eq.~(\ref{eq:map1}) in the weak-lensing regime ($\smash{|\hat{\gamma}|,|\kappa| \ll 1}$) where the reduced shear $\smash{\hat{g} = \hat{\gamma} / \left(1-\kappa\right)}$ is equal to the shear $\hat{\gamma}$:

\begin{equation}
\label{eq:map1}
M_{\rm ap}(\vec{\theta}_0) = \int {\rm d}^2 \vec{\theta} \;Q(|\vec{\theta}-\vec{\theta}_0|) \; \gamma_{\rm t}(\vec{\theta},\vec{\theta}_0),
\end{equation}

\noindent The tangential shear $\gamma_{\rm t}(\vec{\theta},\vec{\theta}_0)$ is computed by replacing $\hat{\epsilon}$ by $\hat{\gamma}$ in Eq.~(\ref{eq:gt}), and is related to the intrinsic galaxy ellipticity $\hat{\epsilon_{\rm S}}$ through $\smash{\hat{\epsilon} = (\hat{\epsilon}_{\rm S} + \hat{g} ) / ( 1+\hat{g}^{*}\hat{\epsilon}_{\rm S} )}$ \citep{Seitz+97}. Throughout this paper, we use the reduced shear, as this corresponds to the quantity observed in data. We also tested our whole pipeline on the shear $\hat{\gamma}$ instead and verified that it does not impact our cosmological forecasts, validating the use of $M_{\rm ap}$ with reduced shear.

As demonstrated in \citet{Schneider+97}, the aperture mass can equivalently be written as a convolution of the convergence $\smash{\kappa(\vec{\theta})}$ and a compensated aperture filter $U(\theta)$:

\begin{equation}
M_{\rm ap}(\vec{\theta}_0) = \int {\rm d}^2 \vec{\theta} \;U(|\vec{\theta}-\vec{\theta}_0|)\;\kappa(\vec{\theta}),
\end{equation}

\begin{equation}
U(\theta) = 2 \int_{\theta}^{\infty} {{\rm d} \theta'\;{Q(\theta')\over \theta'}}  - Q(\theta).
\end{equation}

As $U(\theta)$ is a compensated filter, $\int {\rm d} \theta \, \theta \, U(\theta) = 0$, which means that the residual constant from integrating the shear into the convergence vanishes. This is not the case for classical mass reconstructions \citep[e.g.,][]{KS93} which usually resort to averaging the mass map to zero to lower the impact of that constant, referred to as mass sheet degeneracy. As the integration constant will vary from field to field and between the observational data and the modeled ones, it introduces random variations to the mass levels that could bias the cosmological constraints from mass map statistics. Dealing with the mass sheet degeneracy issue makes $M_{\rm ap}$ a well suited method for cosmology analyses (e.g., peak statistics).

Another important advantage of $M_{\rm ap}$ is that one can define a local noise as the dispersion of $M_{\rm ap}$ assuming no shear, that is only due to galaxy intrinsic ellipticities: 

\begin{equation}
\label{eq:N}
\sigma(M_{\rm ap}(\vec{\theta}_0))=\frac{1}{\sqrt{2}n_{\rm gal}}\left(\sum_{i}{|\hat{\epsilon}(\vec{\theta}_i)|^{2}Q^{2}(|\vec{\theta}_i-\vec{\theta}_0|)}\right)^{1/2}.
\end{equation}

\noindent The $\smash{\sigma(M_{\rm ap}(\vec{\theta}_0))}$ accounts for local variations of the galaxy density that arise from masks or different magnitude depths within a survey.

From Eqs.~(\ref{eq:map}) \&~(\ref{eq:N}), one can derive the local signal-to-noise ratio (S/N), which will be used to measure the height of $M_{\rm ap}$ pixels:

\begin{equation}
\label{eq:SN}
\frac{S}{N} \left( \vec{\theta}_0 \right) = \frac{\sqrt{2} \sum_i Q(|\vec{\theta}_i-\vec{\theta}_0|) \epsilon_{{\rm t}}(\vec{\theta}_i, \vec{\theta}_0)}{\sqrt{\sum_{i}{|\hat{\epsilon}(\vec{\theta}_i)|^{2}Q^{2}(|\vec{\theta}_i-\vec{\theta}_0|)}}}.
\end{equation}

In contrast to \citet{Martinet+18}, we do not include galaxy weights and shear multiplicative biases as this article is primarily intended to assess the potential of $M_{\rm ap}$ map cosmology for future cosmic shear surveys in terms of gain in precision. The impact of the shear measurement systematics, while being an important topic, is beyond the scope of this paper.

The formalism above does not assume any particular function for $Q(\theta)$ besides the circular symmetry. In this work, we use the \citet{Schirmer+07} filter which is a simpler form of an NFW \citep{NFW97} halo mass profile, and is hence optimized to detect dark matter halos associated with small-scale structures in $M_{\rm ap}$ maps,

\begin{equation}
\label{eq:Q}
\begin{aligned}
Q(\theta) =  & \left[1 + \exp \left(6 -150 \frac{\theta}{\theta_{\rm ap}}\right) + \exp \left(-47 +50 \frac{\theta}{\theta_{\rm ap}}\right)\right]^{-1} \\
             & \times \left(\frac{\theta}{x_{\rm c}\theta_{\rm ap}}\right)^{-1} \tanh \left(\frac{\theta}{x_{\rm c}\theta_{\rm ap}}\right),
\end{aligned}
\end{equation}

\noindent where $\theta_{\rm ap}$ is the size of the aperture and $x_{\rm c}$ indicates a change of slope at the angular scale $x_{\rm c}\theta_{\rm ap}$ and is analogous to the $r_{\rm s}$ parameter in the NFW profile. The $x_{\rm c}\theta_{\rm ap}$ is referred to as the effective size of the aperture, and corresponds to the smoothing scale of the mass map in other traditional mass reconstruction \citep[e.g.,][]{KS93}. The first term in Eq.~(\ref{eq:Q}) introduces a smooth transition to 0 in the core and edge of the profile. These cuts avoid the singularity in the aperture center, where $\gamma_{\rm t}$ is not defined, and the contamination from the strong shear regime when centered on a massive halo.

Following \citet{Hetterscheidt+05}, we use $x_{\rm c}=0.15$ as an optimal value for halo detection. Although a particular effective aperture size increases the S/N of halos with the same physical scale in the $M_{\rm ap}$ map, it also acts as a smoothing for shape noise and depends on the galaxy density. \new{We tested six different effective scales ($0.5', 1.0', 1.5', 2.0', 2.5', 3'$) and found that $\theta_{\rm ap} x_{\rm c} = 1.5' $ (corresponding to an aperture size $\theta_{\rm ap} = 10'$) maximizes the cosmological forecasts in the case of our mocks in all tomographic configurations, and therefore only present results for this aperture size. This optimal scale is closely followed by the $1'$ and $2'$ scales in term of forecast precision, in good agreement with other stage IV forecasts (for example \citeauthor{Li+19} \citeyear{Li+19} found an optimal smoothing scale of $2'$ for their LSST mocks). The fact that we find the same optimal $\theta_{\rm ap}$ in the different tomographic cases suggests that the choice of aperture size is dominated by the scale of the matched structures rather than the noise smoothing thanks to the high galaxy density.} We do not explore the combination of $M_{\rm ap}$ statistics measured from multiple smoothing scales as \citet{Martinet+18} showed that this combination of scales only marginally improves the cosmological constraints in the case of $M_{\rm ap}$ map peaks.

\subsection{Fourier space computation}
\label{subsec:Fou}

Aperture masses are conveniently computed in Fourier space, where the convolution becomes a product:

\begin{equation}
   \widetilde{M_{\rm ap}} = - \Re \left[ \; \widetilde{Q \, {\rm e}^{-2{\rm i}\phi}} \; \widetilde{\hat{\gamma}} \; \right],
\end{equation}

\noindent where a tilde denotes Fourier transformed quantities. Carrying out the convolution with FFT significantly reduces the computational time. In our case, for a $1024 \times 1024$ pixel map of $100$ deg$^2$ with a galaxy density of $30$ arcmin$^{-2}$ and a filter scale of $10$ arcmin, the Fourier space approach is $\sim25$ times faster than in real space, dropping from $\sim 50$ to $\sim 2$ minutes on a regular desktop computer.

Although they are equivalent in theory, the real and Fourier space approaches differ in practice: The Fourier approach suffers from a loss of resolution and periodic boundary effects. The latter effect is suppressed by removing a stripe of width $\theta_{\rm ap}$ at the field boundary after the convolution (as is also done in the direct-space computation). While we use exact galaxy positions in real space, we have to bin them on a map before performing the Fourier transform, reducing the resolution to that of the final map.

Despite these limitations, our Fourier-space and real-space approaches agree very well in the absence of masks, producing visually identical maps with a null mean residual per pixel. Exploring the impact on the distribution of S/N values of the $M_{\rm ap}$ map, we find that the Fourier approach introduces a loss of power for the high S/N, with $5$\% fewer pixels with ${\rm S/N}\geq3.0$ compared to the real-space computation. This power leakage due to the discretization of galaxy positions in Fourier space depends on the sampling frequency: it is of $14$\% if we decrease the pixel resolution by a factor two and drops to only $2$\% if we increase the resolution by the same factor (i.e., reconstructing a $512\times512$ and $2048\times2048$ pixels map respectively). We however keep our fiducial resolution ($1024\times 1024$ pixels) to lower the computational time. We also verified that the Fourier reconstruction does not significantly bias the cosmological forecasts. We find almost no difference (below $0.3$\% on any parameter precision and bias) between the forecasts in Fourier and real space, validating the use of the Fourier approach.

In the $M_{\rm ap}$ formalism, observational masks affect all pixels closer to the mask than the aperture size both in the direct and Fourier approaches. It is then mandatory to apply the same mask to the simulations used to build the cosmology dependence of $M_{\rm ap}$. In Appendix~\ref{sec:res;subsec:mask}, we investigate the possible impact of a representative \textit{Euclid} mask on the cosmological forecasts. We find a decrease in the forecast precision consistent with the loss of area due to masking and no bias for both the direct- and Fourier-space map computations. In the rest of the paper we use the Fourier-space approach and do not apply any observational mask.

\section{Building mock catalogs}
\label{sec:sim}

\subsection{SLICS and cosmo-SLICS}

The analysis presented in this work relies on two suites of numerical weak lensing simulations, the Scinet LIght Cones Simulations \citep[][SLICS hereafter]{Harnois-Deraps+15, SLICS18} and the cosmo-SLICS \citep{cosmoSLICS}, which are respectively used for estimating the covariance matrix and modeling the signal of our different statistics. Both are based on a series of 100 deg$^2$ light-cones extracted from $N$-body simulations with the multiple plane technique described in the above references. The underlying gravity-only calculations were carried out by {\sc cubep$^3$m} \citep{cubep3m} and evolve $1536^3$ particles inside a box of comoving side set to $505 ~ h^{-1}\,{\rm Mpc}$. Multiple mass sheets were written to disk at regular comoving intervals, and were subsequently used to generate  convergence ($\kappa$) and shear ($\gamma_{1/2}$) maps for a series of source planes (between 15 and 28, depending on the cosmology) from which the lensing quantities can be interpolated given a position and a redshift.

The SLICS were specifically designed for the estimation of covariance matrices: They consist of $928$ fully independent $N$-body $\Lambda$CDM runs in which the cosmological parameters are fixed to $\Omega_{\rm m} = 0.2905$, $\sigma_8 = 0.826$, $h=0.6898$, $n_{\rm s} = 0.969$ and $\Omega_{\rm b} = 0.0473$, while the random seeds in the initial conditions are varied,  allowing us to estimate covariance matrices with the ``brute force'' ensemble approach. A single light-cone was constructed from each $N$-body run, ensuring a complete independence between the 928 light-cones.

The cosmo-SLICS were run with a similar but complementary $N$-body configuration: The random seeds are fixed, while the cosmological parameters $\Omega_{\rm m}$, $S_8$, $h$, and $w_0$ are sampled at 25 points\footnote{The cosmo-SLICS sample \smash{$S_8=\sigma_8\left(\Omega_{\rm m}/0.3\right)^{0.5}$} instead of $\sigma_8$ since the former probes the $\sigma_8$--$\Omega_{\rm m}$ plane perpendicularly to the lensing degeneracy, and is thus best measured by $\gamma$-2PCF.}. The parameters $n_{\rm s}$ and $\Omega_{\rm b}$ were fixed to the value used in the SLICS. The sampling of this 4D space follows a Latin hypercube, which maximizes the interpolation accuracy. Additionally, at each of these nodes, a pair of simulations was produced, in which the sampling variance was highly suppressed from a careful selection of the initial conditions \citep[see][]{cosmoSLICS} such that any measurement averaged over the pair nearly follows the ensemble mean. In total, ten pseudo-independent light-cones were constructed at every cosmo-SLICS cosmology (five per pair member) by randomly rotating the planes and shifting the origin\footnote{The full suite contains 50 such light-cones, but we verified that our signal converged on the mean with 10.}. A $26^{\rm th}$ cosmo-SLICS simulation was run with the same set-up but with cosmological parameters fixed to the SLICS $\Lambda$CDM model, and is used to mock the observational data, from which we try to estimate the cosmology.

The accuracy of the SLICS and cosmo-SLICS $N$-body simulations was quantified with two-point correlation functions and power spectra in \citet{Harnois-Deraps+15} and  \citet{cosmoSLICS}, and is basically within 2\% of the Cosmic Emulator \citep{Coyote14} up to $k=2.0 ~ h\,{\rm Mpc}^{-1}$, with a gradual degradation of the accuracy at smaller scales.
The covariance matrix measured from the SLICS is also in excellent agreement with the analytical predictions, at least for $\gamma$-2PCF and power spectra, and was shown in \citet{cosmoSLICS} to contain more than 80\% of the ``super sample covariance.'' This is a strong indicator of the high accuracy we  achieve on the covariance matrix estimated with the SLICS.

\subsection{Redshift distribution}

\begin{figure}
    \centering
    \includegraphics[width=0.5\textwidth]{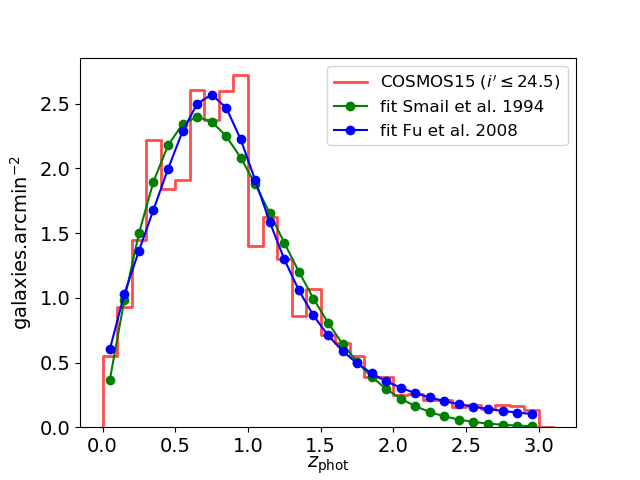}
    \caption{Redshift distribution normalized to $30$ gal.arcmin$^{-2}$. The red histogram corresponds to the COSMOS2015 photometric redshift distribution after a cut at $i'\leq24.5$, the green curve to the fit of a \citet{Smail+94} function to the COSMOS2015 histogram, and the blue curve to a \citet{Fu+08} fit. In this paper we use the \citet{Fu+08} fit that captures the high-redshift tail in contrast to the other function.}
    \label{fig:z}
\end{figure}

From the shear and convergence planes described in the last section, we then constructed a mock galaxy catalog for each light-cone. Because the shear and convergence depend both on the foreground mass and the source redshifts, it is mandatory to reproduce the redshift distribution of the target survey when building the mocks. In order to mimic a survey at the expected depth of the {\it Euclid} Wide Survey \citep{Laureijs+11}, we built a representative redshift distribution from the COSMOS 2015 catalog \citep{Laigle+16} and used a galaxy density of $30$ arcmin$^{-2}$ \citep{Laureijs+11}. We first selected COSMOS galaxies with an $i'$-band magnitude lower than $24.5$ to account for the expected limiting magnitude of the \textit{Euclid} lensing sample in the $VIS$ band, a large filter covering the $r$, $i$, and $z$ bands. We then built the distribution of photometric redshifts in the range $0<z<3$, with a large bin size of 0.1 in redshift in order to smooth the individual structures in the COSMOS field. We fit this redshift distribution with the \citet{Fu+08} function,
\begin{equation}
n(z) = A \frac{z^a+z^{ab}}{z^b+c}.
\end{equation}
We see in Fig.~\ref{fig:z} that this function is able to reproduce the full redshift distribution, including the high-$z$ tail, and smooths isolated peaks that are due to cosmic variance. This would not be the case for the classical \citet{Smail+94} function, $n(z) \propto (z/z_0)^\alpha \exp{[-(z/z_0)^\beta]}$, which fails to capture the high-redshift tail above $z\sim2$. The parameters of the fit that are used to generate the redshift distributions in our simulations are: $A=1.7865$ arcmin$^{-2}$, $a=0.4710$, $b=5.1843$, and $c=0.7259$.

\subsection{Galaxy positions and ellipticities}

Galaxy 2D projected positions are randomly drawn from a uniform distribution. The correlation between the density of galaxies and matter is therefore not captured in our galaxy mocks. Since the average value of the lensing estimator is unaffected by the source galaxy positioning, this choice does not bias our cosmological predictions. However, when using simulations in combination with observations, the positions of galaxies in the simulations are chosen to mimic that of the observation, so as to reduce the impact of masks and shape noise. As a side effect, this positioning introduces wrong correlations between galaxy positions and matter overdensities in the simulations. This is often corrected for through the computation of a boost factor that corresponds to the ratio of the radial number density profiles of galaxies around observed and simulated mass peaks \citep[e.g.,][]{Kacprzak+16,Shan+18}. Although we can neglect this effect in simulation-only analyses, a careful measurement of this bias will be necessary to unlock the full potential of $M_{\rm ap}$ statistics in observational analyses.

When assigning an intrinsic  ellipticity $\hat{\epsilon_{\rm S}}$ to the mock galaxies, each component is drawn from a Gaussian distribution with zero mean and a dispersion $\sigma_\epsilon = 0.26$. This latter value corresponds to Hubble Space Telescope measurements for galaxies of magnitudes $i\sim24.5$ \citep{Schrabback+18}, which will constitute most of the \textit{Euclid} lensing sample and was used previously to investigate the impact of faint
blends on shear estimates \citep[e.g.,][]{Martinet+19}.

Galaxy positions and intrinsic ellipticities constitute the primary source of noise for the extraction of the cosmological signal. To prevent any noise effect on the model of the cosmological dependence of the mass map, we fix the positions, redshifts and intrinsic ellipticities of every galaxy in the cosmo-SLICS mocks for all the different cosmologies, in which only the cosmic shear is changed. These sources of noise are however included in the covariance matrix: Each mass map is computed from an individual realization of the SLICS mocks with different galaxy positions and ellipticities. See Sect.~\ref{sec:cos} for more details.

\section{Data vectors}
\label{sec:vec}

In this section we define several estimators built from the $M_{\rm ap}$ maps (Sect.~\ref{sec:vec:subsec:dv}) as well as the classical $\gamma$-2PCF (Sect.~\ref{sec:vec:subsec:2pcf}), which serves as our baseline comparison method.

\subsection{Sampling the $M_{\rm ap}$ distribution}
\label{sec:vec:subsec:dv}

\begin{figure}
    \centering
\includegraphics[width=0.5\textwidth]{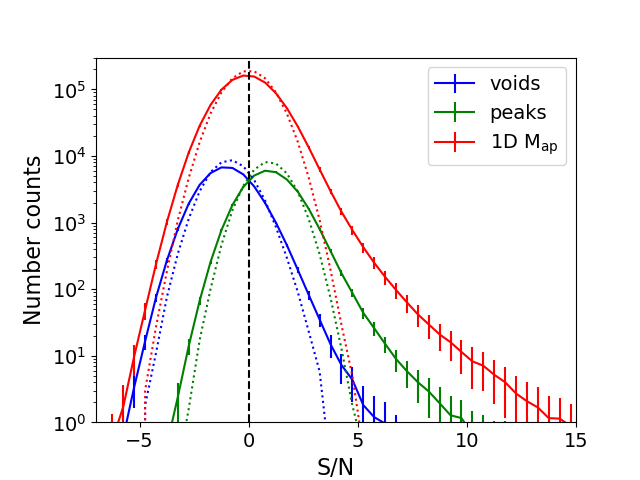}
    \caption{S/N distribution (or data vectors) of three different $M_{\rm ap}$ samplings: voids (blue), peaks (green), and full 1D distribution (red). Dotted lines correspond to the S/N distributions measured in noise-only maps. Numbers are quoted for a $1024\times1024$ pixels map of $100$~deg$^2$ with a filter size $\theta_{\rm ap}=10'$.}
    \label{fig:alldv}
\end{figure}

\begin{figure*}
    \centering
\includegraphics[width=1.0\textwidth]{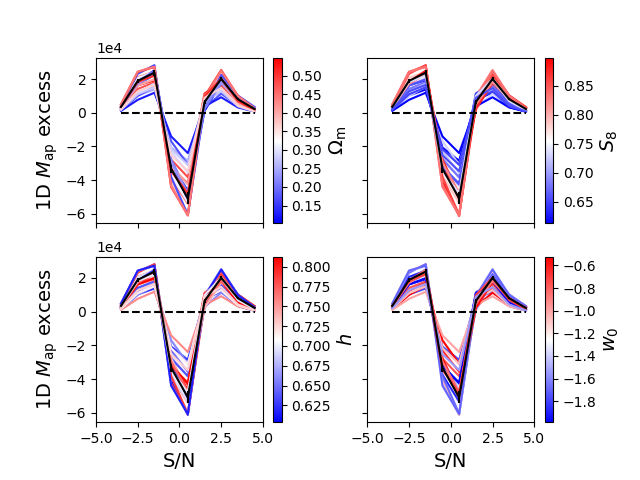}
    \caption{Variation of the excess number of $M_{\rm ap}$ pixels over noise with cosmology. The black curve corresponds to the excess $1$D $M_{\rm ap}$ in the observation mock, with error bars from the diagonal of the covariance matrix. Color-coded are variations in $\Omega_{\rm m}$ (\textit{top left}), $S_8$ (\textit{top right}), $h$ (\textit{bottom left}), and $w_0$ (\textit{bottom right}). A smooth color gradient indicates a correlation between the data vector and the probed cosmological parameter. No tomography has been performed here.}
    \label{fig:dv}
\end{figure*}

Once $M_{\rm ap}$ maps have been generated, there are several choices to sample the 1D distribution of $M_{\rm ap}$. In this article we investigate the use of distributions of $M_{\rm ap}$ pixels that have larger S/N values than their $n$ neighboring pixels. In particular, $n=8$ corresponds to maxima, and $n=0$ to minima. Maximia, or peaks, trace the matter overdensities and in the high-S/N regime are a good representation of massive halos. Peaks therefore share some information with galaxy cluster counts and both probes aim at exploring the non-Gaussian tail of the matter distribution. Peaks, however, present the advantages of being insensitive to the classical selection function and mass-observable relation issues inherent to cluster count studies \citep[e.g.,][]{Sartoris+16, Schrabback+18, Dietrich+19, McClintock+19}. Minima probe the voids of the matter distribution. Alternatively, one can use the full distribution of pixel values (hereafter 1D $M_{\rm ap}$ or lensing PDF).

The number count distribution of these estimators as a function of S/N constitute our data vectors -- DVs for short. They are shown in Fig.~\ref{fig:alldv} for a $100$~deg$^2$ of mock data without tomographic decomposition. 1D $M_{\rm ap}$ is displayed in red, peaks in green, and voids in blue. The curves correspond to the mean over $50$ mock observations from the cosmo-SLICS fiducial cosmology ($2$ different initial conditions for the $N$-body run, five different ray-tracing through the light-cones, and five different shape noise realizations) and the error bars are estimated from the diagonal elements of the covariance matrix computed from the SLICS. As expected from their definition, peaks correspond to higher S/N pixels while the distribution of voids is centered on lower S/N. The dotted lines in Fig.~\ref{fig:alldv} show the DVs measured from noise-only maps computed for the five shape noise realizations we use. We see that $M_{\rm ap}$ maps are dominated by noise, with only a small fraction of the DVs carrying the cosmic shear information. However, the DVs extend to S/N values well above that of the noise-only DVs. These broader wings also imply that the DVs are below the noise near their maximum to  preserve normalization.

In Fig.~\ref{fig:dv}, we exhibit the cosmology dependence by comparing 1D $M_{\rm ap}$ from Fig.~\ref{fig:alldv} (in black) with that from the cosmo-SLICS simulations probing a range of cosmological parameters (blue to red curves). As for Fig.~\ref{fig:alldv}, we show the mean over $50$ $M_{\rm ap}$ maps of $100$~deg$^2$ each, from which we subtracted the noise contribution to better visualize the cosmology dependence. The dependence on $\Omega_{\rm m}$, $S_8$, $h$, and $w_0$ are shown in different panels and colored from blue to red for increasing parameter values. We find a smooth gradient of 1D $M_{\rm ap}$ with $S_8$, and a somewhat smaller variation with $\Omega_{\rm m}$ and $w_0$, highlighting a strong dependence on these parameters. There however does not seem to be any particular dependence on $h$.

In Fig.~\ref{fig:dv} and all the following results we consider only the S/N ranges that can be accurately modeled with our emulator (see Sect.~\ref{sec:cos;subsec:emu}): $-2.5<{\rm S/N}<5.5$ for peaks, $-5<{\rm S/N}<3$ for voids, and $-4<{\rm S/N}<5$ for 1D $M_{\rm ap}$, and use a bin size $\Delta {\rm S/N} = 1$. \new{Fig.~\ref{fig:alldv} suggests that bins of larger S/N could be used in the case of peaks and 1D $M_{\rm ap}$ when no tomography is applied. We find however that these bins only mildly improve the forecast precision (by less than $5$\% and $2$\% for peaks and 1D $M_{\rm ap}$ respectively) as the associated pixel counts are particularly low while they introduce biases due to the large uncertainty of the emulator model in this S/N range. The chosen bin size corresponds to the largest width} for which the cosmological forecasts are not degraded in our analysis. The high correlation between close bins in Fig.~\ref{fig:dv} however tends to show that the information is partly redundant. DV compression, for example through principal component analysis, would probably allow one to reduce the size of the $M_{\rm ap}$ DVs to a few interesting numbers, although we do not explore this possibility in the present article.

\subsection{Shear two-point correlation functions}
\label{sec:vec:subsec:2pcf}

To assess the complementary information from non-Gaussian estimators to traditional ones, we measure the $\gamma$-2PCF in the same mocks on which we compute the aperture mass maps. We use the classical $\xi_+/\xi_-$ definition of $\gamma$-2PCFs \citep{Schneider+02}, which sums the ellipticity correlations over $N_{\rm pairs}$ pairs of galaxies in bins of separation $\theta=|\vec{\theta}_{a}-\vec{\theta}_{b}|$,
\begin{equation}
\label{eq:2pcf}
\xi_\pm \left( \theta \right)=\frac{1}{N_{\rm pairs}} \sum_{a,b} \left[ \epsilon_{\rm t}(\vec{\theta}_{a}) \epsilon_{\rm t}(\vec{\theta}_{b}) \pm \epsilon_{\times}(\vec{\theta}_{a}) \epsilon_{\times}(\vec{\theta}_{b}) \right],
\end{equation}
where $\epsilon_{\rm t}$ and $\epsilon_\times$ correspond to the tangential and cross ellipticity components with respect to the separation vector $\vec{\theta}_a-\vec{\theta}_b$ between the two galaxy positions $\vec{\theta}_{a}$ and $\vec{\theta}_{b}$. As for $M_{\rm ap}$ we do not include observational weights in the former equation.

Although $\xi_+/\xi_-$ can be linked to the integrals of the matter power spectrum weighted by a $0^{\rm th}/4^{\rm th}$ order Bessel function to build a theoretical model of its cosmology dependence \citep[see e.g.,][for a review]{Kilbinger15}, we use the cosmo-SLICS mocks to model the dependence of the $\gamma$-2PCF on cosmology. This ensures a fair comparison between the different probes and also allows for a full combination of DVs when measuring joint constraints. We note that fixing or varying galaxy positions in the cosmo-SLICS mocks does not impact the $\gamma$-2PCF, which are insensitive to source positions.

We measure $\xi_\pm$ with the tree code \textit{Athena} \citep{Kilbinger+14}. We use an opening angle of $1.7$ degrees to merge trees and speed up the computation. This approximation mainly affects the computation at large scales which are not included in our analysis given the size of our simulation mocks ($10\times10$~deg$^2$), and we verified that $\xi_\pm$ do not change for a smaller opening angle of $1.13$ degrees (as advised in the \textit{Athena} readme). We compute the correlations for $10$ separation bins logarithmically spaced between $0.1'$ and $300'$. We note that the use of $N$-body simulations allows us to include scales below $0.5'$; these probe the deeply nonlinear regime of structure formation, for which the dependence on cosmology is difficult to model correctly with theoretical approaches or fit functions. We nevertheless need to cut large scales due to the size of our simulation mocks, and therefore discard the two last bins of $\xi_+$. We also remove the two first bins of $\xi_-$ where the information content is low and which correspond to scales where the matter power spectrum is not fully resolved in the $N$-body simulations. We are left with $8$ bins centered on $(0.15', 0.33', 0.74', 1.65', 3.67', 8.17', 18.20', 40.54')$ for $\xi_+$, and $(0.74', 1.65', 3.67', 8.17', 18.20', 40.54', 90.27', 201.03')$ for $\xi_-$. These scales are similar to that used in current cosmic shear survey \citep[e.g., in KiDS:][]{Hildebrandt+17} with the addition of the small nonlinear scales for $\xi_+$ as they are expected to be used in \textit{Euclid} \citep[also directly calibrated from $N$-body simulations, as in][]{EuclidII}.

\begin{figure*}
    \centering
    \includegraphics{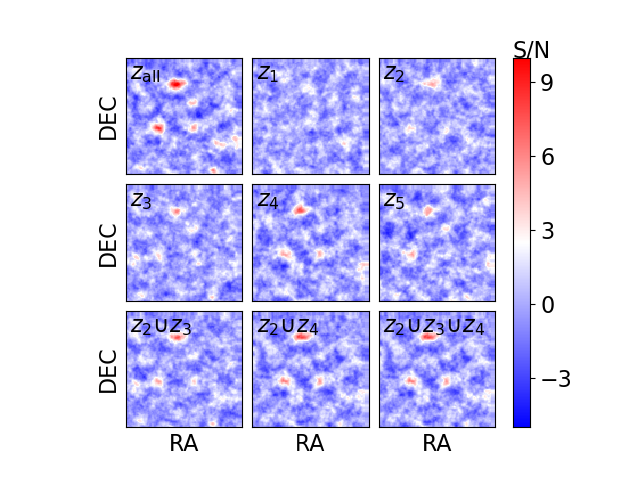}
    \caption{Aperture mass maps of $1$ deg$^2$ obtained from the full redshift distribution ($0<z_{\rm all}<3$), for five continuous redshift slices ($0<z_1<0.4676$, $0.4676<z_2<0.7194$, $0.7194<z_3<0.9625$, $0.9625<z_4<1.3319$, $1.3319<z_5<3$) and for a few cross-combinations of redshift slices ($z_2 \cup z_3$, $z_2 \cup z_4$, $z_2 \cup z_3 \cup z_4$). Massive foreground halos are traced in multiple redshift planes while the noise fluctuates randomly. The cross-$M_{\rm ap}$ terms allow us to retain the information about correlations of structures in between slices.}
    \label{fig:tomomap}
\end{figure*}

\section{Tomography}
\label{sec:vec:subsec:tomo}

Tomography is key to improving the constraints on the dark energy equation of state $w_0$. It is traditionally implemented by slicing the redshift distribution of sources and computing the estimators for the different slices. In principle, a large number of slices allows one to better sample the information of the third dimension. This however decreases the density of galaxies per slice such that measurements become noisier and it is not yet clear what is the ideal trade off between these two effects.

We therefore test various tomographic configurations from $1$ slice (i.e., no tomography) up to ten slices, the target for two-point statistics in {\it Euclid} \citep{Laureijs+11}. The boundaries of each tomographic bin are chosen to have the same galaxy density in each slice. In this way, shape noise affects all slices similarly, which allows us to fairly compare the signal from different redshift bins. All redshifts are assumed to be perfectly known in this article as we do not intend to assess the impact of photometric redshift uncertainties \citep[e.g.,][]{EuclidX} or biases in the mean of the redshift distribution \citep[e.g.,][]{Joudaki+20}.

This tomographic approach to mass maps is however suboptimal as it does not exploit the correlations between different redshift bins. This loss of information likely explains why the $\gamma$-2PCF is better improved by tomography than mass map estimators \citep{Zurcher+20}. Indeed, the former includes correlation functions in between slices (cross terms) in addition to within slices (auto terms). Following the example of $\gamma$-2PCF we develop a new approach in which we explicitly compute cross terms for $M_{\rm ap}$ statistics. This is performed by computing $M_{\rm ap}$ maps for all combinations of redshift slices ($M_{\rm ap}(z_i \cup z_j \cup ...)$) in addition to the auto-$M_{\rm ap}$ terms ($M_{\rm ap}(z_i)$). For example, with $3$ tomographic slices, we use the cross terms $M_{\rm ap}(z_1 \cup z_2)$, $M_{\rm ap}(z_1 \cup z_3)$, $M_{\rm ap}(z_2 \cup z_3)$, $M_{\rm ap}(z_1 \cup z_2 \cup z_3)$, in addition to the usual auto terms $M_{\rm ap}(z_1)$, $M_{\rm ap}(z_2)$, $M_{\rm ap}(z_3)$. The DVs measured in these extra maps cannot be reconstructed from those measured in the $M_{\rm ap}$ maps of the individual slices and contain information about the relative positions of structures in the different redshift slices.

This is illustrated in Fig.~\ref{fig:tomomap} where we show a $1$~deg$^2$ $M_{\rm ap}$ S/N map reconstructed for each individual redshift slice and for a few combinations of redshifts in a five tomographic bin set-up. A striking point is the repetition of high-S/N patterns across slices, which likely originate from the same massive structure. Indeed, a foreground massive halo will distort galaxy shapes in several background source planes and higher-redshift slices contain more information than lower ones. In particular the lowest tomographic slice ($z_1<0.4676$) is almost consistent with a noise-only map. As our DVs are made of pixels counts, the information about the position of structures is lost. The addition of combined redshift bins can compensate (at least partially) for this loss as we include the information about the relative positions of structures probed by the combined redshift planes. We also note that the S/N is larger in the non-tomographic case, which probes five times as many galaxies as a single slice. This $z_{\rm all}$ map exhibits several structures that likely originate from chance superposition of distinct smaller halos \citep{Yang+11} that are often seen in the tomographic slices, and appear as a single large feature in the $z_{\rm all}$ map. The disentangling of these structures with the tomographic approach improves the recovery of the cosmological information embedded in the 3D matter distribution.

For $\gamma$-2PCF, the tomographic DV is defined as the concatenation of the auto-correlations ($\xi_\pm$ in one slice) and cross-correlations ($\xi_\pm$ between pairs of slices). For $M_{\rm ap}$ statistics, we concatenate the DVs measured in every combination of redshift slices, including combinations of more than two slices. The cross-terms significantly increase the size of the DV and with the $928$ realizations of the SLICS simulations used to compute the covariance matrix, the latter becomes too noisy for large number of tomographic slices. For ten tomographic bins, the $\gamma$-2PCF DV would contain $880$ bins requiring several thousand simulations for an accurate estimate of the covariance matrix (and its inverse). While we study the previously used tomography of auto-$M_{\rm ap}$ up to ten slices, we do not include tomography above five slices for the $\gamma$-2PCF and for the cross-$M_{\rm ap}$. This corresponds to DVs of, respectively, $240$ and $279$ elements for a total of $519$ elements in the combined $\gamma$-2PCF and cross-$M_{\rm ap}$ analysis. For larger number of slices, the estimation of the covariance starts to become noisy, especially when combining $\gamma$-2PCF and $M_{\rm ap}$.

\section{Computing cosmological forecasts}
\label{sec:cos}

Forecasts are obtained by finding the cosmological parameters that maximize the likelihood (see Sect.~\ref{sec:cos;subsec:like}) of the model given an observation. Noise is accounted for through the covariance matrix computed from the SLICS simulations (DVs labeled $\vec{x}_{\rm s}$) in Sect.~\ref{sec:cos;subsec:cov}. The model that characterizes the dependence of a given estimator on cosmology is described in Sect.~\ref{sec:cos;subsec:model} and is emulated for any cosmology from the cosmo-SLICS simulations (DVs labeled $\vec{x}_{\rm m}$) in Sect.~\ref{sec:cos;subsec:emu}. The mock observation (DVs labeled $\vec{x}$) consists in one of the cosmo-SLICS model that was not used to calibrate the cosmological dependence.

\subsection{Noise}
\label{sec:cos;subsec:cov}

The noise is accounted for through the covariance matrix, that estimates the correlation between the data vectors $\vec{x}_{{\rm s},i}$ for $N_{\rm s}$ different mock realizations $i$,

\begin{equation}
\Sigma(\vec{\pi}_0) = \frac{1}{N_{\rm s}-1}\sum_{i=1}^{N_{\rm s}}{(\vec{x}_{{\rm s},i}(\vec{\pi}_0)-\bar{\vec{x}}_{\rm s}(\vec{\pi}_0)) \; (\vec{x}_{{\rm s},i}(\vec{\pi}_0)-\bar{\vec{x}}_{\rm s}(\vec{\pi}_0))^{\rm T}}.
\end{equation}

\noindent The $\bar{\vec{x}}_{\rm s}$ denotes the average value of the data vector over its different realizations. We neglect the dependence of the covariance matrix on cosmology and estimate it from the $\Lambda$CDM SLICS simulations at cosmology $\vec{\pi}_0$. This has been shown to be the correct approach in a Gaussian likelihood framework, since varying the cosmology in the covariance can introduce unphysical information \citep{Carron13}.

Each of the $N_{\rm s}=928$ mock realizations is fully independent in terms of all three noise components: sample variance, shape noise, and position noise. We find that the shape noise, introduced by a particular realization of galaxy intrinsic ellipticities, is of the same order as the sample variance. The fact that galaxy positions are random gives rise to ``position noise,'' which adds to the shape noise, but only increases the covariance by a few percents at high S/N and up to $\sim20$\% at S/N of $0$. The position of sources has a weaker impact on large S/N values that correspond to massive halos for which the shear is larger.

\new{Using a finite suite of simulations results in uncertainties in the covariance matrix which propagates into an error on the constraints on the cosmological parameters \citep{Hartlap+07,Dodelson+13,Taylor+14}. This loss of information can be quantified by employing a Student-\textit{t} distribution to describe the distribution of values in each S/N bins \citep{Sellentin+17}. This distribution is the result of the central limit theorem for a finite ensemble of independent realizations, and converges to the Gaussian distribution for an infinite number of samples. The systematically lost information in the variance of each cosmological parameter marginalized over all other parameters can be computed from the number of realizations used in the computation of the covariance matrix ($N_{\rm s}$), the size of the data vector ($N_{\rm d}$) and the number of cosmological parameters inferred ($N_{\rm p}$), via Eq. 42 of \citet{Sellentin+17}. With $N_{\rm s} = 928$, $N_{\rm p}=4$ and  $N_{\rm d}$ varying between 8 bins in the simplest case to 519 bins for the combination of 1D Map and $\gamma$-2PCF with a 5-slice tomography including all cross terms, we find that the accuracy on our errors on each individual parameter is better than $1$\%.}

\begin{figure}
    \centering
\includegraphics[width=0.5\textwidth]{{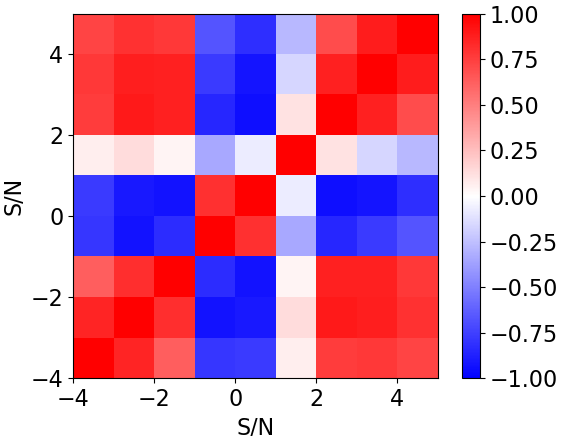}}
    \caption{Correlation matrix computed from 1D $M_{\rm ap}$ in 928 simulations with independent noise realizations (sample variance, shape noise, and position noise).}
    \label{fig:corcov}
\end{figure}

\begin{figure}
    \centering
\includegraphics[width=0.5\textwidth]{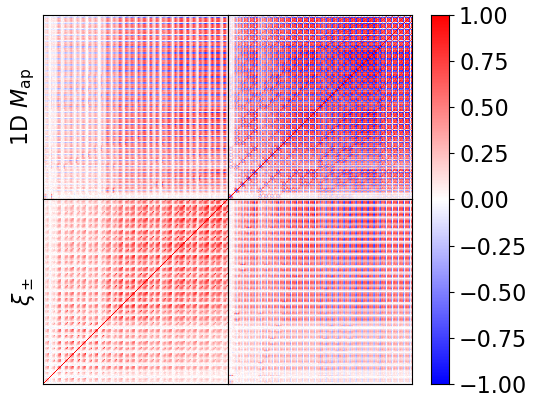}
    \caption{Correlation matrix computed from the combined $\gamma$-2PCF and 1D $M_{\rm ap}$ DVs with five tomographic slices and including cross-terms. $\xi_\pm$ is ordered by increasing redshift from $\xi_\pm(z_1)$, $\xi_\pm(z_1,z_2)$, ..., to $\xi_\pm(z_5)$. 1D $M_{\rm ap}$ is ordered by increasing size of the combination of slices: $M_{\rm ap}(z_i)$, ..., $M_{\rm ap}(z_i \cup z_j)$, ..., $M_{\rm ap}(z_i \cup z_j \cup z_k)$, ..., $M_{\rm ap}(z_i \cup z_j \cup z_k \cup z_l)$, ..., $M_{\rm ap}(z_{\rm all})$. The lower correlations between the two probes indicate a potential gain of cosmological information from their combination.}
    \label{fig:corcovcomb5}
\end{figure}

We show the covariance matrix normalized by its diagonal elements (the correlation matrix) for 1D $M_{\rm ap}$ in Fig.~\ref{fig:corcov}. It displays high correlations between close S/N bins, and presents three different regimes as already seen from the DV excess in Fig.~\ref{fig:dv}. Figure~\ref{fig:corcovcomb5} is similar to Fig.~\ref{fig:corcov} but for a tomographic analysis with $5$ redshift slices and combining with $\gamma$-2PCF. The lower left quadrant of the matrix displays $\xi_\pm$ ordered from low to high redshifts including both auto- and cross-correlations: $\xi_\pm(z_1)$, $\xi_\pm(z_1,z_2)$, ..., $\xi_\pm(z_2)$, $\xi_\pm(z_2,z_3)$, ..., $\xi_\pm(z_5)$. The upper right quadrant shows 1D $M_{\rm ap}$ first for auto-$M_{\rm ap}$, and then by increasing the combination size of cross-$M_{\rm ap}$: $M_{\rm ap}(z_1)$, ..., $M_{\rm ap}(z_5)$, $M_{\rm ap}(z_1 \cup z_2)$, ..., $M_{\rm ap}(z_4 \cup z_5)$, ..., $M_{\rm ap}(z_1 \cup z_2 \cup z_3)$, ..., $M_{\rm ap}(z_1 \cup z_2 \cup z_3 \cup z_4)$, ..., $M_{\rm ap}(z_{\rm all})$. First looking at $M_{\rm ap}$, we note that the non-tomographic correlation pattern is seen in every tomographic auto- and cross-information term. We additionally see correlations in between redshift slices and for combined slices that fade for larger distance between slices. As already noted from the $M_{\rm ap}$ maps (see Fig.~\ref{fig:tomomap}) this is probably due to massive halos being observed from multiple background source planes. The high correlations between cross-$M_{\rm ap}$ terms also suggest that one could find a more compact representation of the tomographic DV. Now focusing on the combination of the two estimators, we find somewhat lower correlations than for individual DVs. This shows that $\gamma$-2PCF and $M_{\rm ap}$ maps indeed probe partially independent information such that their combination should result in a gain of precision in the forecasts.

\subsection{Model}
\label{sec:cos;subsec:model}

The model of the cosmology dependence of each data vector is built from the cosmo-SLICS simulations. For each of the 25 nodes in the parameter space, we compute $M_{\rm ap}$ maps and measure the distribution of S/N of each estimator and then interpolate the DV at any cosmology. In contrast to the computation of the covariance matrix where all sources of noise must be accounted for, we need the model to be as insensitive to noise as possible to ensure the model is accurate in the parameter space close to that of the observational data.

Sample variance is reduced at the level of the simulations and of the ray-tracing. As detailed in \citet{cosmoSLICS}, the cosmo-SLICS $N$-body runs were produced in pairs for which the sample variance is already highly suppressed, allowing us to rapidly approach the ensemble mean by averaging our estimated signal over these pairs. Additionally, ten pseudo-independent light-cones are generated for each of these simulation pairs at every cosmology, using the same random rotations and shifting.  Averaging over these further suppresses the sampling variance associated with the observer's position. To minimize shape noise, galaxy intrinsic ellipticities and positions are kept identical for every cosmology and light-cone, and we generate five different shape noise realizations to lower the noise due to a particular choice of ellipticities. In practice we only change the orientation of galaxies and keep the same ellipticity amplitude, as those should be determined from the observations.
 
This results in a total of 50 mocks of $100$~deg$^2$ per cosmology: (two different initial conditions for the $N$-body simulations) $\times$ (five different ray-tracings) $\times$ (five different shape noise realizations). These numbers are chosen such that adding extra sample variance or shape noise realizations does not affect the cosmological forecasts any more.

\subsection{Emulator}
\label{sec:cos;subsec:emu}

We then emulate the DVs measured from the $25$ nodes at any point in the parameter space through radial basis functions. We use the {\sc scipy.interpolate.rbf} python module that was shown to perform well in interpolating weak-lensing peak distributions \citep[e.g.,][]{Liu+15j,Martinet+18}. We improve from these studies by using a cubic function instead of multiquadric, as we find it to give more accurate results in the case of our mocks. Each bin of S/N of the DV is interpolated independently.

\begin{figure}
    \centering
    \includegraphics[width=0.5\textwidth]{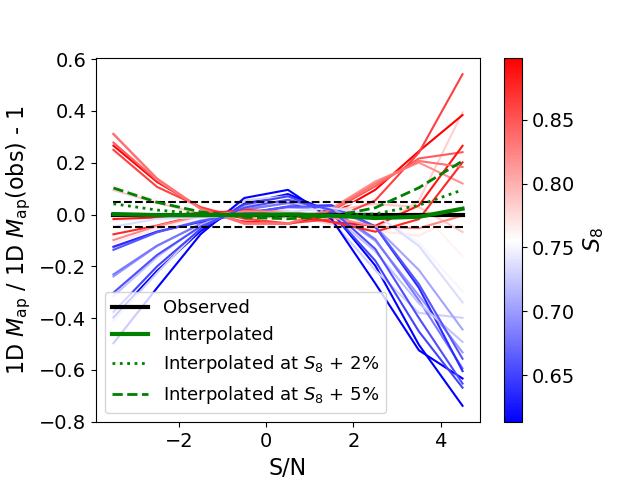} 
    \caption{Testing the emulation of 1D $M_{\rm ap}$. The solid green curve corresponds to the relative excess of the interpolated DV with respect to the measured DV. This error on the interpolation remains below 5\% (dashed black lines) and small in comparison to the relative excess interpolated for a variation of 2\% (doted green curve) and 5\% (dashed green curve) of $S_8$, and to the 25 DVs used to build the emulator (blue to red curves).}
    \label{fig:interp}
\end{figure}

A classical method to verify the accuracy of the interpolation is to remove one node from the cosmological parameter space and to compare the emulation with the measured DV at this node. The proposed verification would however overestimate the errors by measuring them at more than the largest possible distance to a point in the latin hypercube parameter space. We choose instead to evaluate the results of the emulation at the only data point that was not part of the latin hypercube: our observation mock. In Fig.~\ref{fig:interp}, we show the relative error due to the interpolation of 1D $M_{\rm ap}$ at this point (solid green curve). The DV is cut at low and high S/N to retain an accuracy of better than 5\% in the interpolation. We also display the relative difference between the data vector for the 25 values of $S_8$ and the observation (blue to red curves) to compare the cosmology dependence of the model with the error on the interpolation. As can be seen, the error on the interpolation is low compared to the variation due to cosmology. The interpolated DV around the input $S_8$ value (doted and dashed green curves) fits well in the cosmology dependence of the model. Similar figures can be produced for the other 3 cosmological parameters only with a weaker cosmology dependence as already seen from Fig.~\ref{fig:dv}. The stability of the emulator is tested in more details in Appendix~\ref{sec:res;subsec:interp} where we also vary the number of points used in the initial parameter space.

Using the emulator we then generate DVs for a grid of cosmological parameters spanning the full parameter space with $40$ points for each parameter. This might introduce a bias on the measured parameter values of up to the step of the interpolation: $0.011$ for $\Omega_{\rm m}$, $0.007$ for $S_8$, $0.037$ for $w_0$, and $0.005$ for $h$. We find this limitation to be subdominant in the case of the forecasts for $100$~deg$^2$, and close to the $2$\% uncertainties from the cosmo-SLICS simulations. We also apply a Gaussian smoothing to the interpolation results to improve the rendering of the likelihood contours generated for this sparse ensemble of points. In further studies it would be interesting to develop a refining mesh approach or a Markov Chain Monte Carlo sampler to improve the sampling of the parameter space.

\subsection{Likelihood}
\label{sec:cos;subsec:like}

The model built from the cosmo-SLICS simulations allows us to estimate the likelihood: the probability $p(\vec{x}|\vec{\pi})$ to measure a DV $\vec{x}$ for a given set of cosmological parameters $\vec{\pi}=(S_8, w_0, \Omega_{\rm m}, h)$. This likelihood is related to the quantity of interest, the posterior likelihood or the probability of a cosmology given the observed DV $p(\vec{\pi}|\vec{x})$ through Bayes' theorem,
\begin{equation}
p(\vec{\pi}|\vec{x})=\frac{p(\vec{x}|\vec{\pi})p(\vec{\pi})}{p(\vec{x})},
\end{equation}
where $p(\vec{\pi})$ is a uniform prior within the parameter range probed by the cosmo-SLICS simulations, and $p(\vec{x})$ is a normalization factor.

\new{We use a Student-$t$ likelihood, which is a variation to the Gaussian likelihood that accounts for the uncertainty in the inverse of the covariance matrix due to the finite number of simulations used in its estimation \citep{Sellentin+16} that would otherwise artificially improve the precision of the forecasts.} When neglecting the dependence of the covariance matrix on cosmology, this likelihood reads \citep{Sellentin+16}:
\begin{equation}
\label{eq:like}
p(\vec{x}|\vec{\pi}) \propto \left[ 1+\frac{\chi^2\left(\vec{x},\vec{\pi}\right)}{N_{\rm s}-1}\right]^{-N_{\rm s}/2},
\end{equation}
\begin{equation}
\label{eq:chi2}
\chi^2(\vec{x},\vec{\pi}) = (\vec{x}-\vec{x}_{\rm m}(\vec{\pi}))^{\rm T} \; \Sigma^{-1}(\vec{\pi}_0)  \; (\vec{x}-\vec{x}_{\rm m}(\vec{\pi})).
\end{equation}
The $N_{\rm s}=928$ is the number of SLICS simulations used to compute the covariance matrix. 
\new{We verified that every bin of our DVs are compatible with a multivariate Gaussian distribution by comparing their distribution across the SLICS to a Gaussian model drawn from the mean and dispersion of the same bin values in the SLICS. This shows that we have a sufficiently large ensemble of independent realizations that we could use a Gaussian likelihood, but we prefer to apply the Student-$t$ which is more accurate for finite samples (it converges to the Gaussian likelihood for $N_{\rm s} \to \infty$).} $\chi^2(\vec{x},\vec{\pi})$ is a measure of the deviation of a measured DV $\vec{x}$ from a DV $\vec{x}_{\rm m}$ modeled at cosmology $\vec{\pi}$, and accounting for the measurement errors through the precision matrix $\Sigma^{-1}$ at a fixed cosmology $\vec{\pi}_0$.

We note that this Bayesian approach is superior to the Fisher formalism sometimes used in the literature \citep[e.g.,][]{Martinet+15b}, as it does not assume a linear dependence of the DV on the cosmological parameters considered here, an assumption that is most likely invalid in the case of non-Gaussian statistics. Our approach is also closer to an observation-based analysis and allows us to estimate not only the precision on the cosmological parameters but also potential biases.

\begin{figure*}
    \centering
\includegraphics[width=1.0\textwidth]{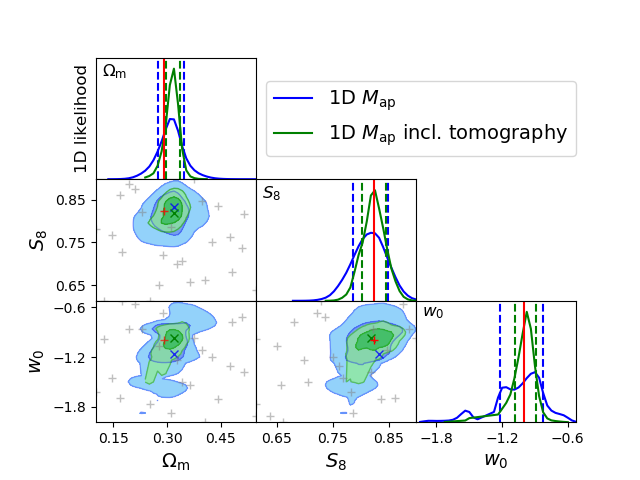}
    \caption{Forecast of cosmological parameters from 1D $M_{\rm ap}$ in a $100$~deg$^2$ survey at \textit{Euclid} depth. Marginalized 2D ($1$ and $2\sigma$ contours) and 1D (full likelihood) constraints are displayed in blue without tomography and in green for a tomographic analysis with $5$ redshift slices and including auto- and cross-$M_{\rm ap}$ terms between redshift slices. Dashed lines correspond to the $1\sigma$ constraints in the 1D marginalized likelihood. The red crosses and lines indicate the input values while the blue and green crosses indicate the best estimate in the 2D constraints. Gray crosses correspond to parameters of the $25$ cosmo-SLICS simulations that are used to estimate the cosmology dependence of 1D $M_{\rm ap}$.}
    \label{fig:forecasts}
\end{figure*}

\subsection{Forecasts}

The cosmological forecasts are obtained by finding the cosmology $\vec{\pi}_{\rm best}$ that maximizes the posterior likelihood. Although we follow a full 4D analysis we display $2$D and $1$D forecasts by marginalizing over the other parameters. The $1\sigma$ and $2\sigma$ contours are calculated as the contours enclosing, respectively, $68$\% and $95$\% of the marginalized likelihood.

Throughout this article, reported forecast values correspond to the $1\sigma$ limit in the $1$D marginalized likelihood. Likewise, the biases are computed as the difference between the best estimate in the $1$D likelihood and the true parameter value in input of the observation mock. Percentage values are in percentage of the true parameter value. We study both the expected precision and bias on $S_8$, $w_0$, and $\Omega_{\rm m}$, but do not present results for $h$. Indeed, our precision forecasts on $h$ are unreliable because the likelihood extends up to the prior limit of that parameter. This advocates for using a wider prior on $h$ when defining the parameter space to run the $N$-body simulations.

Forecasts are computed for a $100$~deg$^2$ survey at \textit{Euclid} depth. Given the few percent accuracy of the cosmo-SLICS simulations, it is not possible to study potential biases on the recovered parameter values for the full $15\,000$ deg$^2$ \textit{Euclid} survey area, since the statistical precision would exceed that of the simulations, and therefore we also refrain from computing accurate forecasts in the latter case. Although our analysis is well suited to explore the potential of $M_{\rm ap}$ map statistics as a cosmological probe, simulations with improved accuracy will be needed for future observations.

\section{$M_{\rm ap}$ statistics forecasts}
\label{sec:res}

We concentrate first on results obtained for the $M_{\rm ap}$ statistics and discuss those from the $\gamma$-2PCF in the next section.

\subsection{1D $M_{\rm ap}$ tomographic forecasts}
\label{sec:res;subsec:tomo}

\begin{table*} 
\caption[]{Cosmological predictions for $100$~deg$^2$ \textit{Euclid}-like mock without tomography and with a five-slice tomography set-up, including cross-$M_{\rm ap}$ terms, for voids, peaks, 1D $M_{\rm ap}$, $\gamma$-2PCF, and for the combination of $M_{\rm ap}$ map estimators with the $\gamma$-2PCF. The ``$\delta$'' values refer to the $1\sigma$ precision forecasts, while ``$\Delta$'' measures the bias between the best fit value and the input. Numbers in parenthesis show the results in percentage of the input value.} 
\centering 
\begin{tabular}{lcccccc} 
\hline 
\hline
 & $\delta S_8$ & $\Delta S_8$ &  $\delta w_0$ & $\Delta w_0$ & $\delta \Omega_{\rm m}$ & $\Delta \Omega_{\rm m}$ \\
\hline
voids no tomo.           & $0.040$ ($4.9$\%) & $-0.020$ ($-2.4$\%) & $0.26$ ($26$\%) & $-0.16$ ($-16$\%) & $0.088$ ($30$\%) & $0.029$ ($10$\%) \\ 
peaks no tomo.           & $0.034$ ($4.1$\%) & $-0.012$ ($-1.5$\%) & $0.29$ ($29$\%) & $0.10$ ($10$\%) & $0.046$ ($16$\%) & $0.017$ ($6$\%) \\ 
1D $M_{\rm ap}$ no tomo. & $0.031$ ($3.8$\%) & $-0.005$ ($-0.6$\%) & $0.20$ ($20$\%) & $0.10$ ($10$\%) & $0.036$ ($12$\%) & $0.017$ ($6$\%) \\ 
$\gamma$-2PCF no tomo.   & $0.041$ ($5.0$\%) & $-0.027$ ($-3.3$\%) & $0.38$ ($38$\%) & $0.10$ ($10$\%) & $0.067$ ($23$\%) & $-0.006$ ($-2$\%) \\ 
\hline 
voids incl. tomo.           & $0.024$ ($2.9$\%) & $0.002$ ($0.3$\%) & $0.12$ ($12$\%) & $0.10$ ($10$\%) & $0.027$ ($9$\%) & $0.029$ ($10$\%) \\ 
peaks incl. tomo.           & $0.022$ ($2.7$\%) & $0.002$ ($0.3$\%) & $0.13$ ($13$\%) & $0.07$ ($7$\%) & $0.024$ ($8$\%) & $0.017$ ($6$\%) \\ 
1D $M_{\rm ap}$ incl. tomo. & $0.021$ ($2.6$\%) & $0.002$ ($0.3$\%) & $0.10$ ($10$\%) & $0.03$ ($3$\%) & $0.019$ ($7$\%) & $0.029$ ($10$\%) \\ 
$\gamma$-2PCF incl. tomo.   & $0.023$ ($2.8$\%) & $-0.005$ ($-0.6$\%) & $0.19$ ($19$\%) & $0.07$ ($8$\%) & $0.034$ ($12$\%) & $0.006$ ($2$\%) \\ 
\hline
\hline
voids + $\gamma$-2PCF no tomo.           & $0.028$ ($3.5$\%) & $-0.005$ ($-0.6$\%) & $0.18$ ($18$\%) & $0.10$ ($10$\%) & $0.046$ ($16$\%) & $0.017$ ($6$\%) \\ 
peaks + $\gamma$-2PCF no tomo.           & $0.027$ ($3.3$\%) & $-0.012$ ($-1.5$\%) & $0.20$ ($20$\%) & $0.10$ ($10$\%) & $0.040$ ($14$\%) & $0.017$ ($6$\%) \\ 
1D $M_{\rm ap}$ + $\gamma$-2PCF no tomo. & $0.027$ ($3.3$\%) & $-0.005$ ($-0.6$\%) & $0.17$ ($17$\%) & $0.10$ ($10$\%) & $0.030$ ($10$\%) & $0.017$ ($6$\%) \\ 
\hline 
voids + $\gamma$-2PCF incl. tomo.           & $0.014$ ($1.7$\%) & $-0.005$ ($-0.6$\%) & $0.08$ ($8$\%) & $0.07$ ($7$\%) & $0.017$ ($6$\%) & $0.017$ ($6$\%) \\ 
peaks + $\gamma$-2PCF incl. tomo.           & $0.015$ ($1.8$\%) & $-0.005$ ($-0.6$\%) & $0.09$ ($9$\%) & $0.07$ ($7$\%) & $0.017$ ($6$\%) & $0.017$ ($6$\%) \\ 
1D $M_{\rm ap}$ + $\gamma$-2PCF incl. tomo. & $0.013$ ($1.5$\%) & $-0.005$ ($-0.6$\%) & $0.06$ ($6$\%) & $0.03$ ($3$\%) & $0.015$ ($5$\%) & $0.029$ ($10$\%) \\ 
\hline
\end{tabular} 
\label{tab:deouf}
\end{table*}

We show in Fig.~\ref{fig:forecasts} and in the upper part of Table~\ref{tab:deouf} the forecasts for a $100$~deg$^2$ \textit{Euclid}-like survey for 1D $M_{\rm ap}$ without tomography (blue contours and curves) and with a tomographic analysis with $5$ redshift slices including both auto- and cross-terms (green contours and curves). We find a good sensitivity of 1D $M_{\rm ap}$ on $S_8$ and $\Omega_{\rm m}$. The sensitivity to $w_0$ is lower but improves more drastically than other parameters when including tomography, as seen especially in the 1D likelihoods. All constraints become tighter in the tomographic case, highlighting the gain of information from probing along the redshift direction. The $1 \sigma$ credible intervals are reduced by including $5$-slice tomography by $32$\% on $S_8$, $49$\% on $w_0$, and $46$\% on $\Omega_{\rm m}$ for $1$D $M_{\rm ap}$. Using only the auto-$M_{\rm ap}$ terms would reduce this gain to $13$\%, $34$\%, and $20$\% on $S_8$, $w_0$, and $\Omega_{\rm m}$, respectively. These results show that cross-$M_{\rm ap}$ terms contain significant additional information to the auto-terms, notably improving $w_0$ forecasts by an extra $\sim50$\%. \new{The reported values assume the same S/N range for the non-tomographic and tomographic cases. We note that with a larger area one might be able to also accurately emulate these statistics for larger S/N bins in the non-tomographic case, such that the gain brought by the tomographic approach could be slightly lower.}

The likelihood contours appear noisy because of the interpolation of the model DVs that loses accuracy when further away from one of the $25$ cosmo-SLICS simulations, represented as gray crosses in Fig.~\ref{fig:forecasts}. We also note a small bias when comparing the best estimates to the true values but it remains within the $1\sigma$ error bars, except for $\Omega_{\rm m}$ when including tomography. Although the precision of the forecasts is greatly enhanced by the tomographic approach, the biases are less affected by it. Only the bias on $w_0$ is significantly reduced with tomography because of the disappearance of the double peak in the 1D likelihood (see Fig.~\ref{fig:forecasts}). The fact that the bias remains constant suggests that it is driven by the simulation parameter space rather than the DV measurements. We explore this further in Appendix~\ref{sec:res;subsec:interp}, where we vary the number of points in the parameter space used to perform the model interpolation. We find that removing up to $5$ points out of the $25$ we use in the fiducial analysis only slightly decreases the forecast precision while the bias increases dramatically. These results confirm the high impact of the number of training nodes on this bias, and demonstrate that a densely sampled parameter space will be necessary when applying $M_{\rm ap}$ statistics to observations.\\

We now investigate how the precision gain varies with the considered number of tomographic slices, and try to identify the optimal number of slices for $M_{\rm ap}$ map cosmological analyses. We show this variation of the forecast precision in Fig.~\ref{fig:tomofor} for $S_8$, $w_0$, and $\Omega_{\rm m}$ for all DVs. The dotted lines correspond to the previous tomography set-up including only the $M_{\rm ap}$ maps of individual redshift slices (auto-$M_{\rm ap}$), and the solid lines to our refined set-up with combinations of redshift slices (auto- and cross-$M_{\rm ap}$). For the former we explore the dependence on up to $10$ redshift slices, and for the latter up to $5$ because of the larger DV size and the corresponding restriction of the calculation of the covariance from SLICS (see Sect.~\ref{sec:cos;subsec:cov}).

When considering only auto-$M_{\rm ap}$, the gain brought by tomography rapidly converges to a fixed value for all three parameters probed. The largest gain is observed when going from $1$ (no tomography) to $2$ slices, and does not evolve any more above $\sim5$ slices. This is in good agreement with results from \citet{Yuan+19} who found for a \textit{Euclid}-like weak-lensing peak analysis that the constraints on $\Omega_{\rm m}$ and $\sigma_8$ are only marginally improved when using $8$ tomographic slices compared to $4$.

This is no longer the case when we additionally include cross-$M_{\rm ap}$ terms. We find that the forecasts are greatly enhanced compared to the previous implementation of mass map tomography, and that the gain keeps increasing up to the largest number of tomographic bins that we could test. In the auto-$M_{\rm ap}$ approach, most of the tomographic gain is likely brought by separating the unlensed foreground galaxies from the background galaxies that carry the cosmic shear information. This is supported by the observation that the gain is mainly concentrated on the low number of slices in this case. In contrast, the cross-terms bring new information by probing correlations between structures across the different redshift slices. This extra information also removes the double-peak in the $w_0$ likelihood, that explains the strange behavior of the auto-$M_{\rm ap}$ constraints between $2$ and $4$ tomographic slices: a small shift in the position of the likelihood maximum with respect to this double peak can significantly reduce the error bars by excluding one of the peaks from the $\pm1\sigma$ uncertainty interval.\\

Our new tomographic approach shows great potential and will allow us to exploit the 3D cosmic shear information with improved constraints at least up to five redshift slices. However, identifying the number of $z$-slices at which the information content saturates will require a larger number of $N$-body simulations in order to improve the accuracy of the model dependence of $M_{\rm ap}$ on cosmology and of the covariance matrix. Finally, we note that the number of $z$-bins that can realistically be employed might also be set by our ability to recover unbiased mean source redshifts in these slices, an effect that is not studied in this article. 

\subsection{Comparison of 1D $M_{\rm ap}$ with peaks and voids}
\label{sec:res;subsec:samp}

\begin{figure}
    \centering
\includegraphics[width=0.48\textwidth]{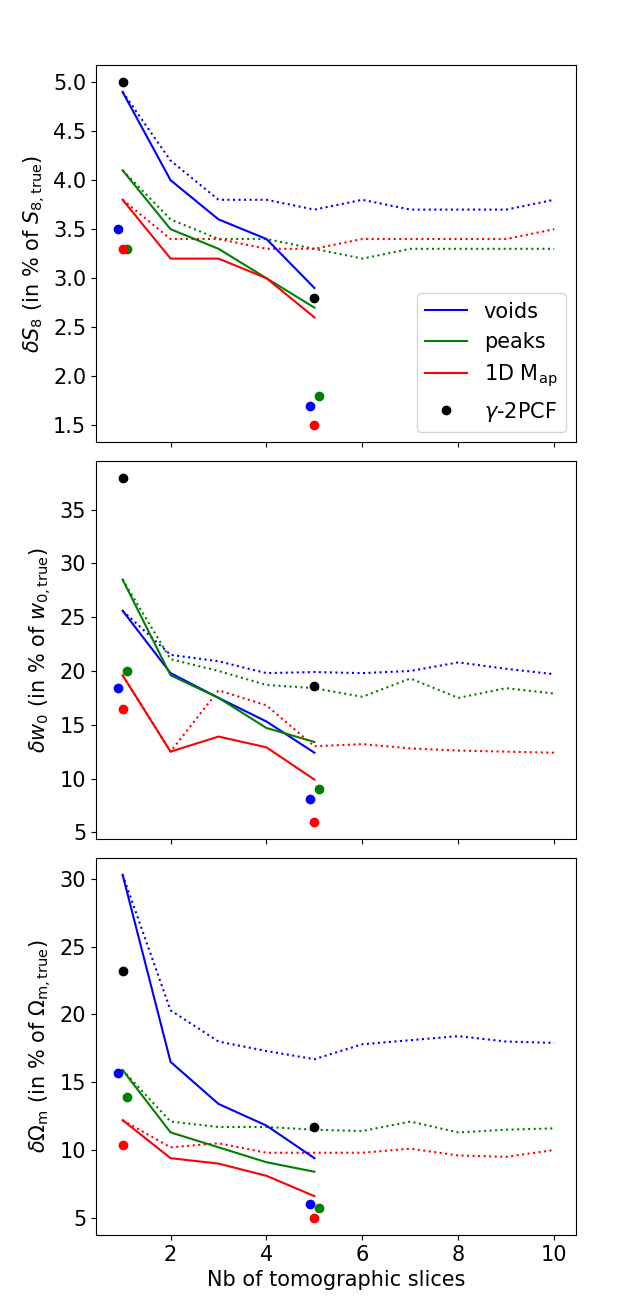}
    \caption{Dependence of the $1\sigma$ error on the number of tomographic bins for $S_8$ (\textit{top}), $w_0$ (\textit{middle}), and $\Omega_{\rm m}$ (\textit{bottom}). Constraints are represented in blue for voids, green for peaks, and red for the full 1D $M_{\rm ap}$ distribution. Dotted and solid lines respectively correspond to the previous tomography set-up (auto-$M_{\rm ap}$ only) and to the improved tomography (auto- and cross-$M_{\rm ap}$). Black dots represent the $\gamma$-2PCF (both auto- and cross-correlations) in configurations with $1$ and $5$ tomographic slices, and color dots the combination between $M_{\rm ap}$ map DV of the same color and $\gamma$-2PCF including all auto- and cross-terms.}
    \label{fig:tomofor}
\end{figure}

In Fig.~\ref{fig:tomofor} and Table~\ref{tab:deouf}, we also present the forecasts for peaks and voids measured in the $M_{\rm ap}$ maps.

$1$D $M_{\rm ap}$ performs better than both peaks and voids regardless of the considered cosmological parameter and tomographic set-up. Peaks show a higher sensitivity to $S_8$ and $\Omega_{\rm m}$ than voids and both estimators perform as well when focusing on $w_0$. This can be explained by the fact that peaks probe matter overdensities, while dark energy also significantly affects the properties of voids. $1$D $M_{\rm ap}$ has a superior constraining power because it includes both peaks and voids. The likelihood is however noisier for $1$D $M_{\rm ap}$ as many $M_{\rm ap}$ pixels contain no signal, but this does not degrade the performance of this estimator. The forecast precision for $1$D $M_{\rm ap}$ is $7$\%, $31$\%, and $23$\% better than for peaks on $S_8$, $w_0$, and $\Omega_{\rm m}$, respectively, without tomography, and $4$\%, $26$\%, and $21$\% when including tomography with five redshift slices. We also note that the relative gain of information from tomography is larger for voids than for peaks or $1$D $M_{\rm ap}$.

With regards to biases in the recovered cosmology, we see that they are quite similar for the different estimators, another hint that they are primarily due to the sampling of the parameter space by the $N$-body simulations, as studied in Appendix~\ref{sec:res;subsec:interp}.

Comparing weak-lensing peak constraints to current surveys, we find a dramatic improvement due to the larger galaxy density and to the fact that higher-redshift galaxies carry more lensing information. With $100$~deg$^2$ at \textit{Euclid} depth without including tomography, we forecast a precision of $4.1$\% on $S_8$, which is almost twice as good as the currently best observational constraints from peak statistics \citep[$7.1$\% using the first $450$~deg$^2$ of KiDS;][]{Martinet+18}. This highlights the potential of forthcoming Stage IV cosmic shear surveys.

Finally, we also experimented with other DVs: the distribution of $M_{\rm ap}$ pixels higher than their $n$ neighbors, for $n$ varying between 0 and 8. $n=4$ is particularly interesting, as it includes (but is not restricted to) the saddle points that can be linked to the presence of filaments \citep{Codis+15}. This estimator behaves very similarly to $1$D $M_{\rm ap}$ but has a lower constraining power. It however exceeds the forecast precision of peaks and voids. The other $n$ values present a smooth degradation of the constraints from $n=4$ to the peaks ($n=8$) and $n=4$ to the voids ($n=0$).

Overall, $1$D $M_{\rm ap}$ performs better than any other estimator. It should then be used to obtain the best possible cosmological constraints from $M_{\rm ap}$ map statistics, unless it is found in the future that it is more severely affected by systematic uncertainties than the other probes.


\section{Combining $M_{\rm ap}$ statistics and shear two-point correlation functions}
\label{sec:res;subsec:2pcf}

\begin{figure*}
    \centering
    \includegraphics{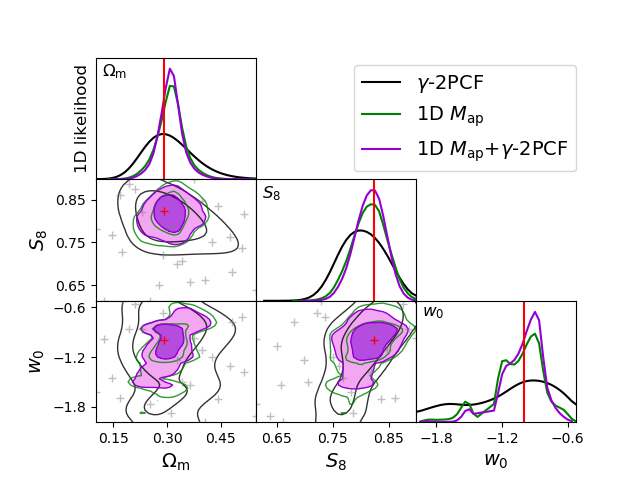}
    \caption{Forecast of cosmological parameters from 1D $M_{\rm ap}$, $\gamma$-2PCF and their combination in a $100$~deg$^2$ \textit{Euclid}-like survey without tomography. Marginalized 2D ($1$ and $2\sigma$ contours) and 1D (full likelihood) constraints are displayed in black for the $\gamma$-2PCF, green for 1D $M_{\rm ap}$, and purple for their combination. The red crosses and lines indicate the truth. Gray crosses correspond to the $25$ simulation points from which the model is built.}
    \label{fig:deouftomo1}
\end{figure*}

\begin{figure*}
    \centering
    \includegraphics{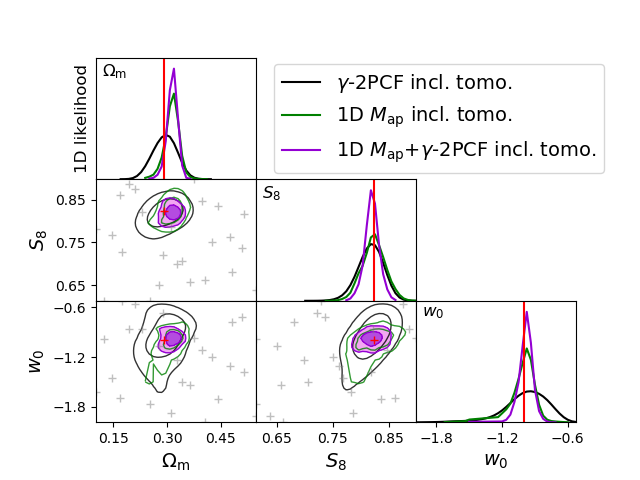}
    \caption{Same as Fig.~\ref{fig:deouftomo1} for a tomographic analysis with five redshift slices, including auto- and cross-$M_{\rm ap}$ terms.}
    \label{fig:deouftomo3}
\end{figure*}


\subsection{Comparison of 1D $M_{\rm ap}$ with $\gamma$-2PCF}

We first compare the statistical power of $M_{\rm ap}$ DVs and $\gamma$-2PCF before combining them. This comparison is based on Fig.~\ref{fig:tomofor} and the values of the top part of Table~\ref{tab:deouf}. When no tomography is included, the $\gamma$-2PCF is outperformed by all $M_{\rm ap}$ DVs for all parameters, except by voids on $\Omega_{\rm m}$. The best estimator, $1$D $M_{\rm ap}$, presents constraints better than the $\gamma$-2PCF by $24$\%, $48$\%, and $47$\% on $S_8$, $w_0$, and $\Omega_{\rm m}$, respectively, highlighting the greater power of non-Gaussian estimators. We also note that the $\gamma$-2PCF presents less noisy likelihoods (see Fig.~\ref{fig:deouftomo1}), as this estimator is less impacted by shape noise than $M_{\rm ap}$ maps.

When applying tomography with $5$ redshift slices and including both auto- and cross-terms, 1D $M_{\rm ap}$ still outperforms the $\gamma$-2PCF for $w_0$ and $\Omega_{\rm m}$, but presents a similar constraining power than the $\gamma$-2PCF on $S_8$. In this set-up, the constraints are still 7\%, 47\% and 44\% better with $1$D $M_{\rm ap}$ than $\gamma$-2PCF on $S_8$, $w_0$ and $\Omega_{\rm m}$. The smaller difference on $S_8$ with the tomographic $\gamma$-2PCF could partially be explained by the favorable definition of this parameter. $S_8$ corresponds to the particular value of $\alpha=0.5$ in \smash{$\Sigma_8 = \sigma_8 \, \left(\Omega_{\rm m}/0.3\right)^\alpha$}, which is chosen to maximize the precision of the $\gamma$-2PCF on that parameter. Since $M_{\rm ap}$ DVs present different degeneracies in the $\sigma_8$--$\Omega_{\rm m}$ plane than the $\gamma$-2PCF \citep[in particular large S/N peaks; e.g.,][]{Shan+18}, they are probably more sensitive to a $\Sigma_8$ defined with a different $\alpha$ parameter.

For $w_0$ and $\Omega_{\rm m}$, we find the $1$D $M_{\rm ap}$ forecasts to be superior to the $\gamma$-2PCF by roughly the same amount without and with tomography. This is in contrast with the previous $M_{\rm ap}$ tomography set-up (doted lines in Fig.~\ref{fig:tomofor}), which is closer to and sometimes even outperformed by the $\gamma$-2PCF for larger number of redshift slices. These results confirm the hypothesis of \citet{Zurcher+20} that the $\gamma$-2PCF benefits more from tomography than mass map estimators in their analysis because they include cross-correlations between redshift slices in the $\gamma$-2PCF but not in their mass reconstruction. In contrast, our addition of the cross-$M_{\rm ap}$ terms allows us to better extract the tomographic information similarly to the $\gamma$-2PCF cross-correlations.

We note here that our constraints on the $\gamma$-2PCF are representative of the \textit{Euclid} survey set-up, and can therefore be safely compared to other estimators in this analysis. Assuming a precision dependence proportional to the square root of the number of galaxies considered, the $18.6$\% constraints on $w_0$ in the five tomography set-up extrapolates to a $1.5$\% forecasts on this parameter for the $15\,000$ deg$^2$ final area, in good agreement with the expected $1$\% precision for $\gamma$-2PCF when including ten tomographic slices \citep{Laureijs+11}. We also verified that $\xi_\pm$ measured in our simulation-based mocks is consistent with theoretical predictions from \citet{Takahashi+12} using the {\sc nicaea} software \citep{Kilbinger+09}. This model is based on halofit \citep{Smith+03} with a calibration on high-resolution $N$-body simulations to account for the nonlinear power spectrum contribution on small scales. The \citet{Takahashi+12} model is within the $1\sigma$ uncertainty of our DV (computed from the covariance matrix) on every scale except on the smallest bin where it is within $2\sigma$.


\subsection{Combination of 1D $M_{\rm ap}$ and $\gamma$-2PCF}

The combination of $M_{\rm ap}$ estimators with the $\gamma$-2PCF is very powerful. It is shown as colored dots in Fig.~\ref{fig:tomofor}. One striking point is the smaller difference between the three $M_{\rm ap}$ DVs forecasts when combined with the $\gamma$-2PCF. This suggests that a large part of the non-Gaussian complementary information to the $\gamma$-2PCF is contained in all $M_{\rm ap}$ DVs.

We also show in Figs.~\ref{fig:deouftomo1} \&~\ref{fig:deouftomo3} the $2$D and $1$D likelihoods for 1D $M_{\rm ap}$ (in green), $\gamma$-2PCF (in black), and their combination (in purple), with one and five tomographic slices. We see a large gain on all parameters when combining estimators. Without tomography, the combined forecasts are largely dominated by the 1D $M_{\rm ap}$ constraints. With tomography, both 1D $M_{\rm ap}$ and $\gamma$-2PCF contribute to the better forecasts, especially for $w_0$.

The quantitative forecasts are displayed in the lower part of Table~\ref{tab:deouf} and can be compared with the upper part values for the individual estimators. We focus here on the combination of the $\gamma$-2PCF with $1$D $M_{\rm ap}$, which is the best $M_{\rm ap}$ estimator. Without tomography the combined forecasts for $S_8$, $w_0$, and $\Omega_{\rm m}$ are, respectively, $34$\%, $56$\%, and $55$\% better than that of the $\gamma$-2PCF alone. The gain is lower when compared to $1$D $M_{\rm ap}$ alone: $13$\% on $S_8$, $16$\% on $w_0$, and $15$\% on $\Omega_{\rm m}$. When applying a $5$ bin tomography with all cross terms, the gain is of $46$\%, $68$\%, and $57$\% on $S_8$, $w_0$, and $\Omega_{\rm m}$, respectively, compared to the $\gamma$-2PCF alone, and $42$\%, $39$\%, and $24$\% compared to $1$D $M_{\rm ap}$ alone. These improved forecasts confirm the good complementarity between non-Gaussian and traditional two-point estimators.


\subsection{Comparison with literature on weak-lensing peak statistics}

We now compare our forecasts with recent results in the literature of weak-lensing peak statistics. Without tomography we find an improvement of $34$\% on $S_8$ when combining peaks and $\gamma$-2PCF compared to the $\gamma$-2PCF alone. This is somewhat higher than the value of $20$\% found in the case of the KiDS data in \citet{Martinet+18}, and could be explained by the higher galaxy density of our Stage IV mocks, which decreases the impact of shape noise in the $M_{\rm ap}$ maps.

When including tomography with auto- and cross-$M_{\rm ap}$, we find an improvement of $36$\% on $S_8$ and $51$\% on $\Omega_{\rm m}$ with peaks and $\gamma$-2PCF compared to $\gamma$-2PCF alone. In order to compare with the literature, we also compute the combined forecasts for auto-$M_{\rm ap}$ only. In that case we find an improvement of $21$\% on $S_8$ and $38$\% on $\Omega_{\rm m}$.

The latter constraints are in very good agreement with recent results from \citet{Li+19} who found an improvement of $32$\% on $\Omega_{\rm m}$ when combining convergence peaks and power spectrum in their tomographic analysis with LSST mocks. These authors also computed forecasts on the amplitude of the primordial matter power spectrum $A_{\rm s}$ (similar to $\sigma_8$) and the sum of the neutrinos that we do not explore in our analysis. Our forecasts on $S_8$ are also close to the $25$\% improvement from the combination of peaks and power spectrum compared to power spectrum alone performed by \citet{Zurcher+20} on DES mocks including tomography.

The similar forecast improvements on $S_8$ and $\Omega_{\rm m}$ between our analysis with auto-$M_{\rm ap}$ terms and the recent literature comforts the robustness of our results. The fact that our constraints are also systematically better when including the cross-$M_{\rm ap}$ terms highlights again the superiority of this new tomographic approach for cosmological analyses with mass maps.

For the first time, we forecast constraints on $w_0$ through a combination of tomographic $M_{\rm ap}$ methods and $\gamma$-2PCF. We find that the combination of peaks and $\gamma$-2PCF improves the constraints on this parameter by $52$\% compared to the $\gamma$-2PCF alone in our tomography set-up (and $37$\% when including only auto-$M_{\rm ap}$). The gain from $M_{\rm ap}$-map peaks is lower than for 1D $M_{\rm ap}$ which also probes the information from other structures. In addition to improve our chances of solving the recent $\sigma_8$--$\Omega_{\rm m}$ tension by increasing the precision on $S_8$, we now know that weak-lensing peak statistics (and more so 1D $M_{\rm ap}$) also probe significant complementary information to the $\gamma$-2PCF on $w_0$, and are therefore a powerful tool to help understand the nature of dark energy from cosmic shear surveys.

\section{Conclusion}
\label{sec:ccl}

In this article, we optimized the method to apply $M_{\rm ap}$ statistics to future cosmic shear surveys (e.g., \textit{Euclid}, LSST, WFIRST). We created \textit{Euclid}-like galaxy mocks of $100$ deg$^2$ based on the SLICS $N$-body simulations to build the covariance matrix, and on $25$ different cosmologies, the cosmo-SLICS simulations, to infer the model dependence of different estimators on cosmological parameters $S_8$, $w_0$, and $\Omega_{\rm m}$. We computed aperture-mass maps from the shear instead of traditional mass-map reconstructions that require the calculation of the convergence first, and verified that realistic observational masks are not biasing the forecasts. We developed a new tomographic approach that includes cross-information between redshift slices in the range $0<z<3$. We emulated the cosmology dependence of three non-Gaussian statistics and of the $\gamma$-2PCF. We forecast their precision and bias on cosmology, maximizing the likelihood between the model, accounting for sample variance and shape noise in the covariance matrix. We then combined these non-Gaussian estimators with the $\gamma$-2PCF, including the correlations between the two probes, to assess the statistical gain. The main results of our analysis are the following:

\begin{itemize}

    \item[$\bullet$] We built different DVs from the $M_{\rm ap}$ maps: the full distribution of pixels ($1$D $M_{\rm ap}$), the local maxima (peaks), and minima (voids). We found that $1$D $M_{\rm ap}$ outperforms other estimators, followed by peaks, and finally voids, both with and without tomography. These estimators are also superior to the $\gamma$-2PCF: in particular, the uncertainties on $w_0$ from the 1D $M_{\rm ap}$ are smaller than for the $\gamma$-2PCF by a factor of $\sim2$, irrespective of the tomographic configuration.\\

    \item[$\bullet$] The explicit computation of the cross-$M_{\rm ap}$ terms from the combinations of redshift slices increases the tomographic forecast precision by an extra $\sim50\%$ compared to the previous tomographic approach which only relies on individual redshift slices (auto-$M_{\rm ap}$ terms in our framework). Only these latter terms have been used in the literature up to now, explaining why authors find, in contrast to us, that non-Gaussian mass-map estimators profit less from tomography than the $\gamma$-2PCF.\\
    
    \item[$\bullet$] The combination of $M_{\rm ap}$ maps with $\gamma$-2PCF significantly improves the forecasts with respect both to $\gamma$-2PCF alone and $M_{\rm ap}$ map alone. This important gain is explained by the different parts of the matter distribution that are probed: mostly Gaussian information from the growth of structure on large scales and at higher redshift from two-point estimators, and the non-Gaussian part from small scales and the lower redshift evolution of structures for mass maps. Uncertainties in $S_8$ are reduced by $\sim 35$\% and $\sim 45$\% (almost twice better), respectively without and with tomography, compared to the $\gamma$-2PCF alone. These numbers are larger than in the recent literature due to our new tomographic approach but are in good agreement when we exclude the cross-$M_{\rm ap}$ terms of our analysis.\\
    
    \item[$\bullet$] For the first time, we also estimated joint tomographic forecasts on $w_0$ finding a decrease in the uncertainty by about a factor of three when combining 1D $M_{\rm ap}$ and $\gamma$-2PCF compared to the $\gamma$-2PCF alone. Further studies are required to better understand how dark energy affects the non-Gaussian part of the matter distribution that is probed by $M_{\rm ap}$ map statistics. The high sensitivity of the halo-mass function to the growth of structures at small scales, or the fact that the $M_{\rm ap}$ statistics studied in this paper are probing a collection of high-order moments could constitute possible ways of explaining this large gain of information compared to two-point statistics.
    
\end{itemize}

These results highlight the feasibility and potential of $M_{\rm ap}$ statistics (including weak-lensing peaks and voids) as complementary probes to the $\gamma$-2PCF in cosmic shear surveys. The large gain on both $S_8$ and $w_0$ offers a greater chance to solve some of the most puzzling current questions in cosmology, such as the recent $\sigma_8$--$\Omega_{\rm m}$ tension and the nature of the accelerated expansion of our Universe. Relying solely on $\gamma$-2PCF to answer these questions would deprive us of a large statistical power already present in our observations, and which is both larger and complementary to that probed by standard two-point estimators inherited from CMB analyses. While the observational systematics of $\gamma$-2PCF are very well understood, combining both estimators in observational data requires us to improve our modeling of the biases inherent to mass map estimators. This will be done by further developing  $N$-body simulations specifically tailored to study the impact of the model emulation, preferentially in a Markov Chain Monte Carlo framework, and by carefully investigating the impact of classical cosmic shear systematics on $M_{\rm ap}$ map estimators: shear multiplicative bias, uncertainty on the mean redshift, baryon feedback, and intrinsic alignments.

\begin{acknowledgements}
We thank Tiago Castro, Carlo Giocoli, \new{the anonymous referee,} and the many KiDS collaborators for useful discussions. NM acknowledges support from a fellowship of the Centre National d'Etudes Spatiales (CNES). JHD acknowledges support from an STFC Ernest Rutherford Fellowship (project reference ST/S004858/1). Computations for the $N$-body simulations were enabled by Compute Ontario (\url{www.computeontario.ca}), Westgrid (\url{www.westgrid.ca}) and Compute Canada (\url{www.computecanada.ca}). The SLICS numerical simulations can be found at \url{http://slics.roe.ac.uk/}, while the cosmo-SLICS can be made available upon request. This work has been carried out thanks to the support of the OCEVU Labex (ANR-11-LABX-0060) and of the Excellence Initiative of Aix-Marseille University - A*MIDEX, part of the French ``Investissements d'Avenir'' program.
\end{acknowledgements}

\bibliographystyle{aa}
\bibliography{spe}

\appendix

\section{Effect of masks}
\label{sec:res;subsec:mask}

\begin{figure*}[!b]
    \centering
    \includegraphics[width=1.0\textwidth]{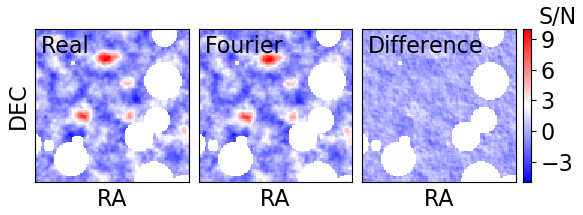}
    \caption{\new{Aperture mass maps of 1~deg$^2$ reconstructed in real (\textit{left}) and Fourier space (\textit{middle}) after applying the star mask, and their difference (\textit{right}).}}
    \label{fig:mask}
\end{figure*}

In this section we explore the impact of observational masks on the forecasts. This is important to verify that masks do not introduce biases when applying $M_{\rm ap}$ statistics to observations.

Similarly to \citet{Laureijs+11}, we build a mask that captures prime features of the expected masks in \textit{Euclid}, as a collection of circular star masks with random positions. We therefore neglect any contribution from noncircular artifacts, for example bad pixels, cosmic rays, and charge transfer inefficiency. Those generally occupy a smaller fraction of the image such that only accounting for stars is a good first order approximation. We draw random radii for the stellar masks from a uniform distribution between $0.29'$ and $8.79'$, corresponding to $0.5$ and $15$ pixels. The largest radius is chosen to match that of the bright Gaia star masks used in the HSC survey \citep{Coupon+18}. This value can be seen as an upper limit as the \textit{Euclid} PSF will be at least four times smaller than that of the Subaru ground-based observation. Using a uniform distribution for star radii does also not reflect the star luminosity function of the Milky Way, overestimating the number of bright stars. With this set-up, we build the mask out of $1000$ random stars, covering about $20$\% of the total $100$ deg$^2$ area of our simulation mocks, and resembling the \textit{Euclid} mask used in \citet{Pires+20}. We use the same mask for all mocks. This is correct for the model determination but slightly underestimates the impact of the mask in the covariance matrix. We recall that our goal is not to apply an exact masking of the data but rather to include the main mask features in order to measure the impact of masking on the $M_{\rm ap}$ map computation. Our mask is presented in Fig.~\ref{fig:mask} for the same $1$~deg$^2$ as in Fig.~\ref{fig:tomomap} and for the real and Fourier space reconstructions. The two reconstructions are visually almost impossible to differentiate one from another\new{, and no remaining structures can be seen in their difference}.


We find that our mask has a negligible impact on the forecasts, both in real and Fourier space, and both without and with tomography. The only difference with the unmasked case is a decrease in the precision consistent with the loss of $20$\% of the probed area.

These results are good news for $M_{\rm ap}$ statistics as they tend to show that we can safely use the Fourier approach to speed up computation without introducing any bias to the cosmological constraints. We however recall that we used several approximations in the mask building and that we also cannot forecast biases for the full coverage of future Stage IV surveys, such that a more complex analysis will be required to verify the possible impact of masks on these surveys to greater precision.

\section{Effect of the cosmological parameters sampling}
\label{sec:res;subsec:interp}

In this section we test the robustness of our forecasts to the number of points explored in the simulated parameter space. This is done by recomputing the forecast when removing the $n$ closest points to the observational mocks in the cosmological parameter space. We stress that this represents a worst-case scenario. We perform this test on $1$D $M_{\rm ap}$ without including tomography. The former case is representative of most bias behaviors in our study as we found the bias to be similar for the different $M_{\rm ap}$ estimators.

We find that removing up to the five closest points to the truth in the studied parameter space weakly affects the forecast precision, while the bias increases for all parameters compared to our fiducial analysis: by $60$\% for $S_8$, $15$\% for $w_0$, and $63$\% for $\Omega_{\rm m}$. Removing more points significantly increases the bias, that becomes larger than the precision and would dominate the forecasts.

These results suggest that the $25$-points modeled by the cosmo-SLICS are sufficient and well adapted to the purpose of this article. However, we cannot find a simple trend of the bias with respect to the number of points included in the parameter space. This prevents us from predicting the necessary number of points to be explored to reach a $1$\% accuracy on $w_0$ from $M_{\rm ap}$ statistics, which will be mandatory to exploit the full potential of future cosmic shear surveys. A more quantitative study with a highly sampled parameter space is required here. In the case of $\gamma$-2PCF, sub-percent level accuracy could be obtained with 250 cosmo-SLICS nodes, as shown in Appendix A of \citet{cosmoSLICS}.

\end{document}